\documentclass[a4paper,oneside,11pt]{article}
\usepackage{cprform}
\usepackage{amsmath,amstext,amsfonts,amsbsy,amssymb,amscd,bbm,epsfig,lscape}

\usepackage{epsfig,macros,cite}
\usepackage{amssymb}
\usepackage{subfigure}

\newcommand{\req}[1]{Eq.~(\ref{#1})}

\newcommand{\rep}[1]{\cite{#1}}
\newcommand{\refig}[1]{Fig.~\ref{#1}}

\newcommand{\ci}[1]{\boldsymbol{#1}}

\newcommand{\Dslash}{\relax{\kern+.25em / \kern-.70em D}}
\newcommand{\Dwd}{D_{\rm\tiny W}}

\newcommand{\Real}{\relax{\mathsf{\Gamma\kern-.35em R}}}
\newcommand{\Int}{\relax{\mathsf{Z\kern-.40em Z}}}

\newcommand{\half}{{\scriptstyle{{1\over 2}}}}

\newcommand{\mren}[1]{m_{{\rm R} #1}}
\newcommand{\muren}[1]{\mu_{{\rm R} #1}}
\newcommand{\obar}[1]{\kern3pt\overline{\kern-2pt #1\kern-0pt}\kern1pt}

\newcommand{\corrbar}[1]{\kern3pt\overline{\kern-2pt #1\kern-0pt}\kern1pt}

\newcommand{\dstwo}{\scriptscriptstyle \Delta {\rm S} =2}

\newcommand{\oVA}[1]{#1_{\rm VA}}

\newcommand{\oVV}[1]{#1_{\rm VV}}
\newcommand{\oAA}[1]{#1_{\rm AA}}

\newcommand{\oVApAV}[1]{#1_{\rm VA+AV}}

\newcommand{\oVApAVren}[1]{\kern3pt\overline{\kern-2pt #1\kern-0pt}\kern1pt_{\rm VA+AV;s}}

\newcommand{\fX}{f_{\rm\scriptscriptstyle X}}
\newcommand{\fP}{f_{\rm\scriptscriptstyle P}}
\newcommand{\fS}{f_{\rm\scriptscriptstyle S}}
\newcommand{\fA}{f_{\rm\scriptscriptstyle A}}
\newcommand{\fV}{f_{\rm\scriptscriptstyle V}}

\newcommand{\FY}{F_{\rm\scriptscriptstyle Y}}
\newcommand{\FVA}{F_{\rm\scriptscriptstyle VA}}

\newcommand{\FVV}{F_{\rm\scriptscriptstyle VV}}
\newcommand{\FAA}{F_{\rm\scriptscriptstyle AA}}

\newcommand{\ZA}{Z_{\rm A}}
\newcommand{\ZV}{Z_{\rm V}}

\newcommand{\ZVApAV}[1]{Z_{\rm VA+AV #1}}

\newcommand{\ZtotVApAV}[1]{\mathcal{Z}_{\rm VA+AV #1}}

\newcommand{\zbar}{\kern3pt\overline{\kern-2pt Z\kern-0pt}\kern1pt}

\newcommand{\zbarVApAV}[1]{\kern3pt\overline{\kern-2pt Z\kern-0pt}\kern1pt_{\rm\scriptscriptstyle VA+AV #1}}
\newcommand{\zrgiVApAV}[1]{\hat Z_{\rm VA+AV #1}}

\newcommand{\mumax}{\mu_{\rm max}}
\newcommand{\mumin}{\mu_{\rm min}}

\newcommand{\icsw}{c_{\rm sw}}

\newcommand{\icA}{c_{\rm\scriptscriptstyle A}}
\newcommand{\icT}{c_{\rm\scriptscriptstyle T}}
\newcommand{\icV}{c_{\rm\scriptscriptstyle V}}
\newcommand{\ibP}{b_{\rm\scriptscriptstyle P}}
\newcommand{\ibS}{b_{\rm\scriptscriptstyle S}}
\newcommand{\ibA}{b_{\rm\scriptscriptstyle A}}
\newcommand{\ibT}{b_{\rm\scriptscriptstyle T}}
\newcommand{\ibV}{b_{\rm\scriptscriptstyle V}}

\newcommand{\ibPtil}{\tilde b_{\rm\scriptscriptstyle P}}
\newcommand{\ibStil}{\tilde b_{\rm\scriptscriptstyle S}}
\newcommand{\ibAtil}{\tilde b_{\rm\scriptscriptstyle A}}
\newcommand{\ibVtil}{\tilde b_{\rm\scriptscriptstyle V}}
\newcommand{\ibTtil}{\tilde b_{\rm\scriptscriptstyle T}}

\begin{document}
\bibliographystyle{mybibstyle}


\begin{titlepage}


\vspace*{-30truemm}
\begin{flushright}
ROM2F/2005/27\\
MS-TP-05-37\\
CERN-PH-TH/2005-268\\
FTUAM-05-20\\
IFT UAM-CSIC/05-55\\
DESY 05-259 \\[10pt]
{\large December 2005}
\end{flushright}
\vspace{9truemm}


\centerline{\LARGE A precise determination of $B_K$ in quenched QCD}
\vskip 9 true mm
\begin{center}
\epsfig{figure=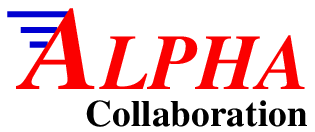, width=22 true mm}\\
\end{center}
\vskip -2 true mm
\centerline{\bigrm  P.~Dimopoulos$^a$, J.~Heitger$^b$, F.~Palombi$^c$,
C.~Pena$^d$, S.~Sint$^e$ and A.~Vladikas$^a$}
\vskip 4 true mm
\centerline{\it $^a$ INFN, Sezione di Roma II}
\centerline{\it and Dipartimento di Fisica, Universit\`a di Roma ``Tor
  Vergata''}
\centerline{\it Via della Ricerca Scientifica 1, I-00133 Rome, Italy}
\vskip 3 true mm
\centerline{\it $^b$ Westf\"alische Wilhelms-Universit\"at M\"unster,
Institut f\"ur Theoretische Physik}
\centerline{\it Wilhelm-Klemm-Strasse 9, D-48149 M\"unster, Germany}
\vskip 3 true mm
\centerline{\it $^c$ DESY, Theory Group, Notkestrasse 85,
D-22603 Hamburg, Germany}
\vskip 3 true mm
\centerline{\it $^d$ CERN, Physics Department, TH Division, 
CH-1211 Geneva 23, Switzerland}
\vskip 3 true mm
\centerline{\it $^e$ Departamento de F\'{\i}sica Te\'orica C-XI and}
\centerline{\it Instituto de F\'{\i}sica Te\'orica C-XVI,}
\centerline{\it Universidad Aut\'onoma de Madrid, Cantoblanco E-28049 Madrid, Spain}
\vskip 10 true mm


\thicktablerule
\vskip 3 true mm
\noindent{\tenbf Abstract}
\vskip 1 true mm
\noindent
{\tenrm The $B_K$ parameter is computed in quenched lattice
QCD with Wilson twisted mass fermions. Two variants
of tmQCD are used; in both of them the relevant $\Delta S = 2$
four-fermion operator is renormalised multiplicatively.
The renormalisation adopted is non-perturbative, with a
Schr\"odinger functional renormalisation condition.
Renormalisation group running is also non-perturbative, up
to very high energy scales. In one of the two tmQCD frameworks the
computations have been performed at the physical $K$-meson mass,
thus eliminating the need of mass extrapolations.
Simulations have been performed at several lattice spacings
and the continuum limit was reached by combining results from
both tmQCD regularisations.
Finite volume effects have been partially checked and
turned out to be small. Exploratory studies have also been
performed with non-degenerate valence flavours.
The final result for the RGI bag parameter, with all sources of
uncertainty (except quenching) under control, is
$\hat B_K =0.789 \pm 0.046$.
}
\vskip 3 true mm
\thicktablerule
\vspace{10truemm}

\eject
\end{titlepage}

\section{Introduction}
\label{sec:intro}

Indirect CP-violation in $K \rightarrow \pi \pi$ decays is expressed
in terms of the $\epsilon$-parameter. Its known experimental value,
combined with theoretical input from neutral $K$-meson oscillations,
defines a hyperbola in the complex plane of the unitarity triangle
(for a review see ref.~\cite{Buras:CPrev}). The theoretical prediction of the
oscillation amplitude is obtained in an operator product expansion
framework, as the product of a single Wilson coefficient (known
to NLO in perturbation theory \cite{rm1:4ferm-nlo,mu:4ferm-nlo})
and the matrix element
$\langle \bar K^0 \vert O^{\Delta S = 2} \vert K^0 \rangle$
of a dimension-6, $\Delta S = 2$ operator
\begin{eqnarray}
O^{\Delta S = 2} &\equiv& [\bar s \gamma_\mu (1-\gamma_5) d] \,\,
[\bar s \gamma_\mu (1-\gamma_5) d] \nonumber \\
&=& O_{\rm VV+AA} - O_{\rm VA+AV} \, .
\end{eqnarray}
Here $s$ and $d$ are strange and down quark fields.
Within the square brackets spin and colour indices are
saturated. In obvious notation, the operators  $O_{\rm VV+AA}$ and
$O_{\rm VA+AV}$ are the parity-even and -odd parts of
$O^{\Delta S = 2}$. For historical and technical reasons,
this matrix element is usually normalised by its
value in the vacuum saturation approximation (VSA). It is thus 
expressed in terms of the $B_K$ parameter:
\begin{eqnarray}
B_K \equiv \dfrac{\langle \bar K^0 \vert O_{\rm VV+AA} \vert K^0 \rangle}
{\langle \bar K^0 \vert O_{\rm VV+AA} \vert K^0 \rangle_{VSA}}
= \dfrac{\langle \bar K^0 \vert O_{\rm VV+AA} \vert K^0 \rangle}
{\dfrac{8}{3} F_K^2 m_K^2} \, ,
\end{eqnarray}
with  $m_K$ the $K$-meson mass and $F_K$ its decay constant.
The above expression involves the matrix element of the
parity-even operator $O_{\rm VV+AA}$ only. 
The parity-odd matrix element
$\langle \bar K^0 \vert O_{\rm VA+AV} \vert K^0 \rangle_{VSA}$
vanishes identically in QCD.

$B_K$ is an intrinsically non-perturbative quantity, which can be
computed in the lattice regularisation of QCD. The results of the
various recent $B_K$ measurements have been summarised in
ref.~\cite{Dawson:lat05} (see also references in this review).
Besides quenching, which is a still uncontrolled
source of systematic error, the most important source of
uncertainty in these computations arises from the operator
renormalisation. In schemes which respect chiral symmetry,
the operator $O_{\rm VV+AA}$ is multiplicatively
renormalisable. Thus, the corresponding physical matrix
element can in principle be accurately obtained by a series
of non-perturbative (lattice) measurements of the bare matrix
element and the operator renormalisation constant, at several
values of the lattice spacing, followed by a continuum limit
extrapolation. This is the case of Ginsparg-Wilson fermions,
where chiral symmetry-breaking effects are negligible.
With Wilson fermions, the breaking of chiral symmetry induces
mixing with four other dimension-6
operators~\cite{Marti:4fPT,Bernardetal:4fPT,Bernard:lat87,
Guptaetal:bk,Doninetal:4fRIMOM} in the parity-even sector:
\begin{eqnarray}
(O_{\rm R})_{\rm VV+AA} = Z_{\rm VV+AA}(g_0, a\mu) \Big [ O_{\rm VV+AA} (g_0)
+ \sum_{i = 1}^4 \Delta_i(g_0) O_i(g_0) \Big ] \,.
\end{eqnarray}
The operators $ O_i(g_0)$ belong to different chiral
representations than $O_{\rm VV+AA}$. The mixing coefficients
$\Delta_i(g_0)$ are finite functions of the bare coupling,
while the renormalisation constant $Z_{\rm VV+AA}(g_0, a\mu)$
diverges logarithmically. This mixing pattern is the reason
behind the relatively poor precision of $B_K$, measured with
Wilson fermions.

Two proposals have attempted to resolve this issue. 
They are both based on the 
observation~\cite{Bernard:lat87,Doninetal:4fRIMOM}
that, even in the absence of chiral symmetry, the parity-odd
operator $O_{\rm VA+AV}$ is protected from mixing with others,
by discrete symmetries:
\begin{eqnarray}
(O_{\rm R})_{\rm VA+AV} = Z_{\rm VA+AV}(g_0, a\mu) O_{\rm VA+AV} (g_0) \,.
\end{eqnarray}
The first proposal~\cite{BK:WI} consists in obtaining the physical
$K^0 - \bar K^0$ matrix element of $O_{\rm VV+AA}$ from
a correlation function of the renormalised operator
$O_{\rm VA+AV}$, through axial Ward identities. The method has
been put to test in ref.~\cite{SPQR:bk}. The $B_K$ estimate of this
method turned out to be compatible with the result of
the standard computation
(with operator subtractions). Unfortunately, the 
correlation function of $O_{\rm VA+AV}$, being the product of four
composite operators, turned out to be noisy and the error
was as large as the one of the standard computation. Thus,
this method is successful in eliminating an important source
of systematic errors (operator subtraction) at the cost of
increased statistical fluctuations.

In this work we implement the second proposal~\cite{tmqcd:pap1},
which is based on the addition of a so-called
``twisted mass term'' to the Wilson fermion action.
This entails loss of parity and partial loss of flavour
symmetry at finite lattice spacing, recovered in the continuum limit. 
On the other hand, some renormalisation properties are greatly 
ameliorated in the twisted mass QCD (tmQCD) formalism. 
The relevant case for us is the renormalised
$\langle \bar K^0 \vert O_{\rm VV+AA} \vert K^0 \rangle$
matrix element, which in the tmQCD formalism may be expressed
in terms of the parity-odd operator $O_{\rm VA+AV}$.
The tmQCD action differs from the standard one by an additional
soft term, which does not modify renormalisation properties
in mass independent renormalisation schemes. In particular, 
the operator $O_{\rm VA+AV}$ remains multiplicatively renormalisable,
with the same renormalisation constant and running as with
Wilson fermions. Thus finite subtractions are avoided
in the tmQCD determination of $B_K$.

With respect to previous studies, we have introduced several
new features:
\begin{itemize}
\item The bare matrix elements have been obtained in two distinct
lattice tmQCD formulations. In the first formulation, the down
quark is twisted, with a twist angle $\taa = \pi/2$, while
the strange quark is a standard (untwisted) Wilson
fermion.
In the second formulation we have a twisted flavour doublet of
down and strange flavours, with twist angle $\taa = \pi/4$.
Having two independent $B_K$ estimates allows a better control
of the continuum limit extrapolations. We have implemented
Schr\"odinger functional (SF) boundary conditions for these
computations.
\item Quenched simulations with standard Wilson fermions are
carried out at heavier quark masses, due to the presence of
exceptional configurations. Since our tmQCD variant with
$\taa = \pi/4$ is free of exceptional configurations,
$B_K$ can be computed at the physical $K$-meson mass value.
Thus the error due to extrapolations from higher masses
is eliminated.
\item In ref.~\cite{ssf:vapav} the operator $O_{\rm VA+AV}$ has been
renormalised non-perturbatively in several SF renormalisation
schemes and its Renormalisation Group (RG) running has been
computed non-perturbatively from low energies up to
scales of several tens of GeV. We have used these results
in order to obtain the Renormalisation Group Invariant (RGI)
counterpart of $B_K$. Our result is essentially free of
any uncertainty due to Perturbation Theory (PT).
\item Although our main results have been obtained with
degenerate down and strange quarks, we have also performed some
exploratory studies with non-degenerate flavours. Also,
some simulations have been performed in order to probe finite
volume effects.
\end{itemize}
Our result confirms earlier ones, obtained from other lattice
regularisations (domain wall, overlap, staggered) albeit with a major
control of the various sources of error. An interesting discrepancy
with the result of the most recent detailed study of $B_K$ with Wilson
fermions (see ref.~\cite{SPQR:bk}) is also resolved.

The paper is organized as follows: In sect.~\ref{sec:tmQCD} we
review the tmQCD basics, emphasising the cases of interest. We show
how, by choosing either a twist angle $\taa = \pi/2$ or $\taa=\pi/4$,
we can map the operator $O_{\rm VV+AA}$ onto $O_{\rm VA+AV}$, thus achieving
multiplicative renormalisation without subtractions. 
In sect.~\ref{sec:sfhme} we
review how four-fermion operator matrix elements can be obtained
from SF correlation functions. In sects.~\ref{sec:res_pi2} and
\ref{sec:res_pi4} we present our raw results for tmQCD with
$\taa = \pi/2$ and $\taa = \pi/4$ respectively.  In
sect.~\ref{sec:bkcon} we compute the RGI-$B_K$, extrapolated to the
continuum limit. Several more technical aspects are discussed
in the appendices.

Preliminary results of our work have appeared in
refs.~\cite{lat03:proc_BK,lat04:proc_BK}. Based on a similar approach,
the $\bar B^0 - B^0$ matrix element, with a static bottom flavour, is
currently under study~\cite{BBarstat:latt04,BBarstat:latt05}.

\section{Twisted mass QCD and weak matrix elements}
\label{sec:tmQCD}

Twisted mass QCD has been designed to eliminate
exceptional configurations in (partially) quenched 
lattice simulations with light Wilson quarks~\rep{tmqcd:pap1}.
In its original formulation, it describes a
mass-degenerate isospin doublet $\psi$ of Wilson quarks for which, 
besides the standard mass term, a so-called twisted mass term
$i\mu_{\rm q}\bar{\psi}\gamma_5\tau^3\psi$ is introduced. 
The properties of tmQCD have been studied in
detail in~~\rep{tmqcd:pap1}, where, in particular, 
its equivalence to standard two-flavour QCD has been established.
We discuss here two extensions of this framework, which are
suited to the extraction of $B_K$.

\subsection{Twisted mass QCD with twist angle $\pi/2$}
\label{subsec:actpi2}

The first variant has already been mentioned in ref.~\rep{tmqcd:pap1},
and corresponds to a Euclidean continuum action of the form
\begin{equation}
 \label{tmQCD_action2}
 S_F = \int{\rm d}^4 x \,\, [ \bar{\psi}(x) (\Dslash 
+ m_l + i\mu_l \gamma_5 \tau^3 )\psi(x) +
\bar s(x) (\Dslash + m_s ) s(x) ]\ .
\end{equation}
Here, the two light flavours form an isospin doublet
$\psi^T = (u,d)$, $\tau^3$ is a Pauli matrix and
$\mu_l$ and $m_l$ are the twisted and standard
quark mass parameters. 
For the light quark sector the physical quark mass is given 
\begin{eqnarray}
M_l = \sqrt{m_l^2 + \mu_l^2}\ ,
 \label{quark_mass_M}
\end{eqnarray}
and the twist angle is defined by
\begin{equation}
 \tan \taa = \dfrac{\mu_l}{m_l}\ .
 \label{eq:tan}
\end{equation}
The considerable advantages of the choice $\taa = \pi/2$, for which
$m_l =0$ and $M_l = \mu_l$, will be discussed below.
We will refer to this case as ``fully twisted'', since the
physical quark mass $M_l$ is determined by the
twisted mass parameter alone. 

The equivalence of this theory to standard QCD, established
in~~\rep{tmqcd:pap1}, is based on axial transformations
of the quark fields and corresponding spurionic transformations of the
mass parameters. Let us relabel the fields and masses of the above
theory by $\psi^{(\taa)}, m_l^{(\taa)}$ and $\mu_l^{(\taa)}$. The
corresponding quantities in standard QCD (which is tmQCD at
$\taa = 0$) are given by $\psi^{(0)}, m_l^{(0)}$ and $\mu_l^{(0)}=0$.
The two theories are then related by the axial field transformations
\be
 \psi^{(\taa)} \to \psi^{(0)} = R(\taa) \psi^{(\taa)}, \qquad
 \psibar^{(\taa)} \to \psibar^{(0)} = \psibar^{(\taa)} R(\taa)\ ,
 \label{axial_transf} 
\ee
and the transformation of the mass parameter
\be
  m_l^{(0)} = m_l^{(\taa)} \cos(\taa)  + \mu_l^{(\taa)} \sin(\taa) \ ,
 \label{mass_transf}
\ee
where
\begin{equation}
  R(\taa) = 
\exp\left\{\frac{i}{2}\gamma_5 \taa\tau^3 \right\} \ .
 \label{rotations}
\end{equation}
In tmQCD we denote Euclidean correlation functions by
\be
\left\langle O[\psi,\psibar]\right\rangle_{(\alpha)} = 
{\cal Z}^{-1}\int_{\rm fields} O[\psi,\psibar]\rme^{-S},
\ee
where $O[\psi,\psibar]$ denotes some
multilocal gauge invariant field. The fermionic part of 
the action is given in~(\ref{tmQCD_action2}). The
relation between standard QCD and tmQCD correlation
functions, are then expressed as
\begin{eqnarray}
   \left\langle O\left[\psi,\psibar\right]
   \right\rangle_{(\alpha)}
 &=& \left\langle O\left[R(-\alpha)\psi,\psibar R(-\alpha)\right]
   \right\rangle_{(0)},
  \label{identity_1}\\[5pt]
   \left\langle O\left[\psi,\psibar\right]
   \right\rangle_{(0)}
 &=& \left\langle O\left[R(\alpha)\psi,\psibar R(\alpha)\right]
   \right\rangle_{(\alpha)}.
 \label{identity_2}
\end{eqnarray}
Hence, a given correlation function in tmQCD with twist
angle $\alpha$ is interpreted as the linear
combination on the r.h.s.~of \req{identity_1}. If instead we
are given a standard QCD correlation function, then
\req{identity_2} tells us how it is represented in tmQCD at
twist angle $\alpha$.

In the formal continuum framework, all these relations between
tmQCD and standard QCD quantities are readily obtained from
the chiral transformations~(\ref{axial_transf}). Their extension to
the lattice regularised theory with Wilson fermions, which is the
lattice regularisation of choice in the present work, is more intricate.
As shown in~\rep{tmqcd:pap1}, the same relations are realised 
between the renormalised quantum field theories, provided 
the renormalisation procedure is set up with some care. Moreover, the
definitions (\ref{quark_mass_M}) and (\ref{eq:tan}) are understood to
hold for suitably renormalised masses $M_{{\rm R},l}$, $m_{{\rm R},l}$ 
and $\mu_{{\rm R},l}$, such that the 
twist angle $\taa$ is free of renormalisation.

\subsubsection{Relations between composite fields}

The axial transformations~(\ref{axial_transf}) induce a mapping
between composite fields. For the quark bilinear operators
\be
\label{bilinears}
 S_{ij} =  \bar{\psi}_{i} \psi_{j}, \quad
 P_{ij} =  \bar{\psi}_{i} \gamma_5 \psi_{j}, \quad
 A_{\mu,ij} =  \bar{\psi}_{i} \gamma_\mu\gamma_5 \psi_{j}, \quad
 V_{\mu,ij} =  \bar{\psi}_{i} \gamma_\mu \psi_{j},
\ee
we have
\begin{align}
  S_{sd}^{(0)} &= \cos\left(\tfrac{\taa}{2}\right) S_{sd}^{(\taa)}
  \hskip 4.8pt  
  - i \sin\left(\tfrac{\taa}{2}\right) P_{sd}^{(\taa)}, 
 \label{eq:Sus_transf}  \\[5pt]
  P_{sd}^{(0)} &= \cos\left(\tfrac{\taa}{2}\right) P_{sd}^{(\taa)}
  \hskip 3.2pt
  - i \sin\left(\tfrac{\taa}{2}\right) S_{sd}^{(\taa)},
 \label{eq:Pus_transf}  \\[5pt]
  A_{\mu,sd}^{(0)} &= \cos\left(\tfrac{\taa}{2}\right)
  A_{\mu,sd}^{(\taa)}
  - i \sin\left(\tfrac{\taa}{2}\right) V_{\mu,sd}^{(\taa)},   \\[5pt]
  V_{\mu,sd}^{(0)} &= \cos\left(\tfrac{\taa}{2}\right)
  V_{\mu,sd}^{(\taa)} \hskip 1.8pt
  - i \sin\left(\tfrac{\taa}{2}\right) A_{\mu,sd}^{(\taa)} \ .
 \label{eq:transform_ops3} 
\end{align}
For the four-quark operators under consideration we have:
\begin{eqnarray}
 \label{Opm_twist_1}
 O^{(0)}_{\rm VV+AA} &=& 
 \cos\left(\taa\right)O^{(\taa)}_{\rm VV+AA}
  -i\sin\left(\taa \right)O^{(\taa)}_{\rm VA+AV}\ , \\[5pt]
 \label{Opm_twist_2} 
  O^{(0)}_{\rm VA+AV} &=& 
 \cos\left(\taa \right)O^{(\taa)}_{\rm VA+AV}
 -i\sin\left(\taa \right)O^{(\taa)}_{\rm VV+AA}.
\end{eqnarray}
The case of interest is $\alpha=\pi/2$, for which
we have
\begin{align}
\label{eq:tmA2}
  A^{(0)}_{\mu,sd} &= \dfrac{\sqrt{2}}{2} [ A_{\mu,sd}^{(\pi/2)} 
  - i V_{\mu,sd}^{(\pi/2)} ]\ ,  & \\[5pt]
\label{eq:tmVA2}
 O^{(0)}_{\rm VV+AA} &= - i O^{(\pi/2)}_{\rm VA+AV}, & 
 O^{(0)}_{\rm VA+AV} = - i O^{(\pi/2)}_{\rm VV+AA} \ .
\end{align}
The above formal expressions imply the following relation between
renormalised operator matrix elements:
\begin{eqnarray}
\langle K^0 \vert \,\, (O_{\rm R})^{(0)}_{\rm VV+AA} \,\, \vert \bar K^0 \rangle =
- i \langle  K^0 \vert \,\, (O_{\rm R})_{\rm VA+AV}^{(\pi/2)} \,\, \vert \bar
K^0 \rangle \,.
\end{eqnarray}
The advantage of computing $B_K$ from the matrix element on the rhs
(i.e. by performing a tmQCD computation), is that  $O_{\rm VA+AV}$
is multiplicatively renormalisable, whereas the
matrix element on the lhs involves $O_{\rm VV+AA}$, which requires additive
renormalisation.

\subsection{Twisted mass QCD with twist angle $\pi/4$}
\label{subsec:actpi4}

We also implemented a second variant of tmQCD, in which the down
and strange quarks are grouped into a  flavour doublet
$\psi^T = (s,d)$:
\begin{equation}
 \label{tmQCD_action4}
 S_F = \int{\rm d}^4x \,\, \left[ \bar{\psi}(x) (\Dslash 
+ m_l + i\gamma_5 \tau^3 \mu_l )\psi(x) +
\bar u(x) (\Dslash + m_u ) u(x) \right]\ .
\end{equation}
This formulation of the theory is suitable for $B_K$ studies with degenerate
light and strange valence flavours. 
In the quenched approximation, the
first term alone is adequate for $B_K$ simulations, since
only strange and down valence quarks are involved. In this
approximation it is straightforward to also accommodate non-degenerate
valence flavours, by introducing diagonal $2 \times 2$ mass
matrices in place of $m_l$ and $\mu_l$ (see ref.~\cite{tmqcd:DIrule})
for details\footnote{In the present work, simulations with
non-degenerate valence quarks have only been carried out with
twist angle $\alpha = \pi/2$.}).

The general relationships of subsect.~\ref{subsec:actpi2},
defining total quark mass, twist angle and chiral rotations
between fermion fields, mass parameters and correlation functions, 
hold also in this case. Once more, they are valid both formally and
between renormalised quantities.

\subsubsection{Relations between composite fields}

In the present tmQCD formulation, the chiral rotations between
composite fields become
\begin{eqnarray}
  S_{sd}^{(0)} &=& S_{sd}^{(\taa)} ,
 \label{eq:Sus_transf4} \\[5pt]
  P_{sd}^{(0)} &=& P_{sd}^{(\taa)} ,
 \label{eq:Pus_transf4} \\[5pt]
  A_{\mu,sd}^{(0)} &=& \cos\left(\taa\right) A_{\mu,sd}^{(\taa)}
  - i \sin\left(\taa\right) V_{\mu,sd}^{(\taa)},  \\[5pt]
  V_{\mu,sd}^{(0)} &=& \cos\left(\taa\right) V_{\mu,sd}^{(\taa)}
  - i \sin\left(\taa\right) A_{\mu,sd}^{(\taa)}.
 \label{eq:transform_ops34} 
\end{eqnarray}
For the four-quark operators of interest we have:
\begin{eqnarray}
 \label{Opm_twist_3}
 O^{(0)}_{\rm VV+AA} &=& 
 \cos\left(2\taa\right)O^{(\taa)}_{\rm VV+AA}
  - i\sin\left(2\taa \right)O^{(\taa)}_{\rm VA+AV}, \\[5pt]
 \label{Opm_twist_4} 
  O^{(0)}_{\rm VA+AV} &=& 
 \cos\left(2\taa \right)O^{(\taa)}_{\rm VA+AV}
  - i\sin\left(2\taa \right)O^{(\taa)}_{\rm VV+AA}.
\end{eqnarray}
At $\alpha=\pi/4$ we have
\begin{align}
\label{eq:tmA4}
  A^{(0)}_{\mu,sd} &= \dfrac{\sqrt{2}}{2} [ A_{\mu,sd}^{(\pi/4)} 
  - i V_{\mu,sd}^{(\pi/4)} ],  & \\[5pt]
 O^{(0)}_{\rm VV+AA} &= - i O^{(\pi/4)}_{\rm VA+AV}\ , &
 O^{(0)}_{\rm VA+AV} = - i O^{(\pi/4)}_{\rm VV+AA}\ .
\label{eq:tmVA4}
\end{align}
The formal above expressions imply for the renormalised WME of interest
\begin{eqnarray}
\langle K^0 \vert \,\, (O_R)_{\rm VV+AA}^{(0)} \,\, \vert \bar K^0 \rangle =
- i \langle  K^0 \vert \,\, (O_R)_{\rm VA+AV}^{(\pi/4)} \,\, \vert
\bar K^0 \rangle \ .
\end{eqnarray}
Here, as in the $\pi/2$ case, we see that the QCD four-fermion WME of
interest is mapped onto a tmQCD WME which involves the
multiplicatively renormalisable four-fermion operator $O_{\rm VA+AV}$.
So from the renormalisation point of view, the $\pi/4$ version is
equally advantageous.

\subsection{Flavour symmetry, strangeness and mass degeneracy}

The two tmQCD formalisms exposed above have distinct characteristics,
which merit some discussion. The $\pi/2$ case refers to a light
flavour doublet, while the strange quark is regularised in the
standard way. Thus simulations may be naturally performed with
non-degenerate down and strange flavours.
In particular, simulations may get close to the physical
situation, since the twisted light quarks 
are protected from exceptional configurations, 
whereas the non-twisted strange quark is heavy 
enough to remain unaffected by this problem.

At this point we note that most quenched computations of $B_K$ have
remained with mass degenerate down and strange quarks,
simply because quenched chiral perturbation theory
indicates the appearance of potentially dangerous quenched chiral logs
as soon as one departs from this situation~\cite{sharpe:Qchlogs,rev:sharpe,SharpeZhang,GoltLeung1}.
In this unphysical situation, the problem 
of exceptional configurations
re-appears for the strange quark. Thus one is forced to stay with 
fairly massive pseudoscalar mesons of about $600~\MeV$, 
where the problem of exceptional configurations is
believed to be absent, at least with a non-perturbatively $\Oa$ improved
action~\cite{mbar:pap3}. In quenched computations with degenerate
strange and down masses, one must therefore compute $B_K$ 
with the $K-$meson tuned to several values
above its physical mass and then extrapolate to the physical point.

While in the $\alpha=\pi/2$ scenario some tuning of the quark mass parameters
is necessary to achieve mass degeneracy between the down and
the strange quark (see below), this unphysical situation is naturally
obtained in the $\pi/4$ case where down and strange 
quarks form a flavour doublet. 
The problem of exceptional configurations is 
now absent. This enables the computation of $B_K$ with degenerate 
valence quarks tuned so as to have the $K$-meson at its physical
mass value, avoiding extrapolations from heavier masses.

More generally, it must be realised that partial loss of flavour
symmetry at finite lattice spacing is the price one has to pay
for the attractive features of tmQCD. Surprisingly large flavour breaking effects have been observed with                                  
maximally twisted Wilson quarks at $\csw=0$ in refs.~\cite{Jansen:2005cg,Farchioni:2005hf}. 
We have also looked at flavour breaking effects and found 
that these are reasonably small and rapidly decreasing towards the 
continuum limit (cf. sect.~\ref{sec:res_pi2}). The question whether these different 
findings are due to differences of the lattice action, parameters or the 
choice of observables will be the subject of further study.

\subsection{Improvement considerations}

Lattice quantities are computed at several fixed UV cutoffs and, after
they are renormalised, they are extrapolated to the continuum limit.
Symanzik improvement, if applied, ensures a better control of the
extrapolating procedure. It involves adding $\Oa$ counterterms
both in the lattice action and the operators. In this work we will be
using the Clover improved action~\cite{impr:SW} and improved
currents (in the spirit of ref.~\cite{impr:lett,impr:pap1}).

Use of the Clover improved action implies that our meson mass
estimates are subject to finite spacing effects which are $O(a^2)$.
In order to remove $\Oa$ cutoff effects from $B_K$, we would also
need to improve the relevant matrix elements of the four-fermion 
operator and the axial current in the tmQCD regularisation.
The improvement pattern of dimension-3 quark bilinear operators 
in the quenched approximation is given in Appendix~\ref{sec:appimp}.
For tmQCD with untwisted (strange) and twisted (down)
quarks, these results give for the currents
\begin{eqnarray}
  (A_{\rm R})_{\mu,sd} &=& \ZA [1 + \frac{1}{2} \ibA a m_{q,s} ]
  [ A_{\mu,sd} + c_{\rm A} a \tilde \partial_\mu P_{sd}
    - i\frac{1}{2}  a \mu_l \ibAtil V_{\mu,sd} ]\,,
\label{eq:impA2} \\[5pt]
  (V_{\rm R})_{\mu,sd} &=& \ZV [1 + \frac{1}{2} \ibV a m_{q,s} ]
  [ V_{\mu,sd} + c_{\rm V} a \tilde \partial_\nu  T_{\mu\nu,sd}
    - i \frac{1}{2} a \mu_l \ibVtil A_{\mu,sd} ]\,,
\label{eq:impV2}
\end{eqnarray}
where the subscript ${\rm R}$ refers to renormalised quantities
and $\tilde \partial$ stands for the symmetric lattice derivative.
These renormalised currents combine as in \req{eq:tmA2} to give
the improved axial current in the $\alpha = \pi/2$ case.
For tmQCD with the down and strange quarks in the same (twisted)
doublet, the results of \cite{tmqcd:pap2} can be directly taken over
(with the up quark replaced by the strange one). We thus have
\begin{eqnarray}
  (A_{\rm R})_{\mu,sd} &=& \ZA [1 + \ibA a m_{\rm q} ]
  [ A_{\mu,sd} + c_{\rm A} a \tilde \partial_\mu  P_{sd}
    - i a \mu_l \ibAtil V_{\mu,sd} ] \,,
\label{eq:impA4} \\[12pt]
  (V_{\rm R})_{\mu,sd} &=& \ZV [1 + \ibV a m_{\rm q} ]
  [ V_{\mu,sd} + c_{\rm V} a \tilde \partial_\nu  T_{\mu\nu,sd}
    - i a \mu_l \ibVtil A_{\mu,sd} ] \,.
\label{eq:impV4}
\end{eqnarray}
These renormalised currents combine as in \req{eq:tmA4} to give
the improved axial current in the $\alpha = \pi/4$ case.
The renormalisation constants and improvement coefficients used here
may be found in Appendix~\ref{sec:appZ}.

The improvement of four-fermion operators is a far more difficult
procedure. Though feasible in principle, it is rendered impractical
by the proliferation of counterterms. We will hence not proceed in
this direction.\footnote{The proposal of
ref.~\cite{FrezzoRoss2} is an interesting alternative, which will
also not be pursued here.}

Following these considerations, we have always used the Clover action
in our simulations. The four-fermion operator is left
unimproved. The implications of current improvement to the O($a$)
effects of $B_K$ require further discussion. We will repeatedly
revert to this issue at later stages.

\subsection{Tuning of quark masses}
\label{sec:masstune}

With the continuum actions~(\ref{tmQCD_action2}) 
and~(\ref{tmQCD_action4}) the twist angle is
directly determined by the ratio of the
mass parameters $m_l$, $\mu_l$ in the action.
This is no longer the case once tmQCD
has been regularised with Wilson type quarks.
Rather, the tuning of the twist angle $\taa$ to the
preferred values $\pi/2$ and $\pi/4$ requires 
the implementation of~\req{eq:tan} for renormalised mass parameters. 
We discuss this issue in some detail.

We denote the subtracted bare quark mass for Wilson fermions
by $a m_{\rm q} = 1/(2 \kappa) - 1/(2 \kappa_{cr})$, $\kappa$ 
being the hopping parameter. Whenever we need to identify the
quark flavour $f$, we will denote the corresponding quantities by
$a m_{{\rm q},f}$ and $\kappa_f$. For the untwisted strange quark in the
$\pi/2$ formulation, the renormalised quenched quark mass is given by
\begin{eqnarray}
\mren{,s} &=& Z_{\rm m} [ m_{{\rm q},s} ( 1 + b_{\rm m} a m_{{\rm q},s} ) ] \,,
\label{eq:msren}
\end{eqnarray}
while the light quark masses renormalise as follows:
\begin{eqnarray}
\mren{,l} &=& Z_{\rm m} [ m_{{\rm q},l} ( 1 + b_{\rm m} a m_{{\rm q},l} ) 
+ \tilde b_{\rm m} a \mu_l^2] \,,
\nonumber
\\[10pt]
\muren{,l} &=& Z_\mu \mu_l ( 1 + b_\mu a m_{{\rm q},l} ) \,.
\label{eq:mlren}
\end{eqnarray}
The above expressions are valid up to $O(a^2)$ corrections. For the
strange flavour, \req{eq:msren} is the standard relationship
between renormalised quark mass and $\kappa_s$. For the light sector,
fixing the twist angle to $\pi/2 + O(a^2)$ amounts to setting
$\mren{,l} = 0$ with $\muren{,l}$ positive.
In other words, once a value $a \mu_l$ is
chosen, we tune $\kappa_l$ so as to satisfy
\begin{equation}
a  m_{{\rm q},l} = - \tilde b_{\rm m} ( a  \mu_l )^2 \,.
\label{qmasspi2}
\end{equation}

The $\pi/4$ case is somewhat less trivial. In order to ensure that 
the twist angle acquires this value to $O(a^2)$, we must
impose $\mu_{\rm R} = m_{\rm R}$ to this order in the cutoff effects
(flavour indices are suppressed, as the formalism applies to
degenerate down and strange quarks). In terms of eqs.~(\ref{eq:mlren})
this means
\begin{equation}
a m_{\rm q} = \dfrac{1}{Z \ZA} a \mu_{\rm q} \big \{ 1 +
\big [ \dfrac{1}{Z \ZA} (b_\mu - b_{\rm m} ) - Z \ZA \tilde 
b_{\rm m} \big ]
a \mu_q \big \} \, ,
\end{equation} 
with $Z \equiv Z_{\rm m} / (Z_\mu \ZA)$. For a given choice
of $a \mu_{\rm q}$ and $\kappa_{cr}$, $\kappa$ is tuned so that 
$a m_{\rm q}$ satisfies the above relation.

In Appendix~\ref{sec:appZ} we collect the known results for the
renormalisation constants and improvement coefficients required above.

\section{SF Correlation functions at large time separations \label{s_corr}}
\label{sec:sfhme}

We now derive explicit expressions for the representation
of Schr\"odinger functional correlation functions in terms of
intermediate physical states. For quark bilinear dimension-3
operators (e.g. the pseudoscalar density, or the axial vector current)
this has been discussed in ref.~\cite{mbar:pap2}. Here we
recapitulate the derivation and generalise it, in a straightforward
manner, to the four-quark dimension-6 operator of interest.
We then discuss how this formalism extends to the case of tmQCD.

\subsection{Quantum mechanical representation of the \SF}

The QCD \SF is defined as the QCD partition function in Euclidean
space-time, with quark and gluon fields obeying periodic boundary
conditions in space (with period $L$) and
Dirichlet boundary conditions in time at the hypersurfaces 
$x_0=0$ and $x_0=T$. We assume homogeneous boundary conditions, 
i.e.~the spatial components of the gauge potentials and 
the Dirichlet components of the quark
and anti-quark fields are taken to vanish at the two time boundaries. 
The Schr\"odinger functional can then be written as~\cite{SF:LNWW,SF:stefan1}
\bes
 {\pf} = \langle {\rm i_0}| \rme^{-T \ham } \projector |{\rm i_0}\rangle
 \label{e_Z}\, ,
\ees
where $\initial$ is a state which is implicitly defined 
by the Dirichlet conditions, and carries the quantum numbers 
of the vacuum state. The presence of $\projector$
implies a projection  onto the subspace of gauge invariant states,
and  $\ham$ denotes the Hamilton operator of QCD formulated on a torus
of volume $L^3$. It is the existence of the Hamiltonian operator
which allows for the definition of a time-zero quantum mechanical
Hilbert space and the corresponding operator representation of
Euclidean correlation functions. On the lattice, $a\,\ham$ is defined as 
the negative logarithm of the transfer matrix and 
it is Hermitian provided the transfer matrix
itself is Hermitian and positive. This is indeed the case with 
the standard Wilson gauge action and unimproved Wilson quarks with both 
standard and twisted mass terms~\cite{Luscher:TM,tmqcd:pap2}.
In principle, this property is lost in the presence of O($a$) 
improvement terms in the action. However, 
the ensuing unitarity violations are usually considered harmless 
since they occur close to the cutoff scale and thus do not affect the physics
at low energies~\cite{Luscher:1984is}.

The quantum mechanical representation of correlation functions
then proceeds via the introduction of a set
of (gauge invariant) eigenstates of $\ham$,
\bes
 &&|n,q\rangle\,,   \quad n=0,1,\ldots \;, \\[10pt]
 && \ham \, |n,q\rangle = E^{(q)}_{n} |n,q\rangle \,,
\ees
with normalisation
$
\langle n',q'|n,q\rangle = 
   \delta_{n,n'} \, \delta_{q,q'}
$.
Here, $q$ is a multi-index comprising all quantum numbers
corresponding to the exact lattice symmetries, and
$n$ enumerates the energy levels in the channel specified by $q$.
We do not indicate the momentum of the states $|n,q\rangle$, since 
we will always be working with correlation functions that project states with 
vanishing (spatial) momentum.

Note that in general the set of conserved lattice quantum numbers $q$
is smaller than in the continuum, due to the explicit breaking
of symmetries by the lattice regularisation.
Standard Wilson quarks have the nice property to conserve
an exact SU($\Nf$) flavour symmetry, besides
parity and charge conjugation symmetries. The only
difference to the continuum classification of particle states then
consists in the breaking of rotational symmetries by
the lattice regularisation, which implies that angular momentum
is not a good quantum number in general. However, this mainly 
affects higher spin states, and is irrelevant for the pseudoscalar
meson states at hand.
In the following we first give an account of 
this continuum like situation with standard Wilson 
quarks before turning to the more complicated case of tmQCD.

\subsection{Specific cases of SF correlation functions}

For the calculation of $B_K$, we are interested in specific
correlation functions of gauge invariant dimension-3 quark bilinear
operators $X$ (where $X$ may denote scalar and pseudoscalar densities
$S$, $P$ or currents $V_\mu$, $A_\mu$), the four-fermion operator
$\oVA{O}^{\dstwo}$ and the dimensionless boundary quark fields
\bes
 \op{ds} = {{a^{6}}\over{L^3}} \sum_{\vecy, \vecz}
           \zetabar_{d}(\vecy)\gamma_5\zeta_{s}(\vecz) 
        \,, \quad
 \op{ds}' = {{a^{6}}\over{L^3}} \sum_{\vecy, \vecz}
           \zetabarprime_{d}(\vecy)\gamma_5
                            \zetaprime_{s}(\vecz)\,,
                            \label{e_op}
\ees
which have been discussed in~\cite{impr:pap1}.
In terms of these quantities we define the gauge invariant
correlation functions
\bes
  \fX(x_0)   =  - \dfrac{L^3}{2}\langle
                 X_{sd}(x) \, \op{ds}
         \rangle \label{e_fa} \,,
\ees
where $\langle \dots \rangle$  denotes the usual path integral average.
Upon specifying $X=S,P,V_\mu, A_\mu$ we denote the above correlation as
$\fX = \fS, \fP, \fV, \fA$ respectively. Similarly, for a $\Delta S =2$ 
four-fermion operator $Y$ we define the correlation function
\bes
\FY(x_0) = L^6 \langle  \op{ds}' \,\,  Y (x)  \,\, \op{ds}  \rangle\,.
\label{eq:fVA}
\ees
Upon specifying $Y=\oVA{O}, \oVV{O},\oAA{O}$ we denote the above correlation as
$\FY = \FVA, \FVV, \FAA$ respectively. Finally, in order to 
obtain properly normalised hadronic states one needs to 
consider the boundary-to-boundary
correlation function
\bes
  f_1 & = & -\frac12
         \langle \op{sd}' \, \op{ds} \rangle \label{e_f1}\,.
\ees
Clearly, in the standard lattice formulation
the parity-odd correlation functions $\fS, \fV$ and $\FVA$
vanish identically, but this will not be the case in tmQCD,
where the corresponding fields will be re-interpreted
according to the discussion in sect.~2.

The correlation functions $\fX$ have the quantum mechanical
representation~\cite{mbar:pap2}.
\bes
  \fX(x_0) = \pf^{-1}\,\dfrac{L^3}{2}\,\initialt 
                   \rme^{-(T-x_0)\ham} 
                   \opX  \rme^{-x_0 \ham }
                  \projector \initialK \,, \quad
                  a \leq x_0 \leq T-a \,,\label{e_fX_qu}
\ees
where $\opX$ is the corresponding time-independent operator, 
and the state $\initialK$ has the quantum numbers of the K-meson
with momentum zero.
Analogously for the correlation function $\FY = \FVV, \FAA$ we obtain
\bes
  \FY(x_0) = \pf^{-1}\, L^6 \, \initialKt  \rme^{-(T-x_0)\ham} 
                   \opY  \rme^{-x_0 \ham }
                  \projector \initialK \,, \quad
                  a \leq x_0 \leq T-a \,,\label{e_fY_qu}
\ees
i.e. the operators $\op{ds}$ and $\op{ds}'$ 
carry the quantum numbers of a $K^0$ by construction.
We also have
\bes
  f_{1} = \pf^{-1}\frac12 \,\initialKt \rme^{-T \ham } \projector
                   \initialK  \label{e_f1_qu} \, .
\ees
The asymptotic behavior of $\fX(x_0)$ (with $X=A_0, P$), $\FY(x_0)$
(with $Y = O_{\rm VV+AA}$) and $f_1$ for large
values of both $x_0$ and $T-x_0$ (with $L$ unspecified at this stage)
is as follows
\bes
  \fX(x_0) &\approx& 
     \frac{L^3}{2}\, \rho \, \langle 0,0| \opX | 0,K\rangle   
                       \, \rme^{-x_0 m_K } \times
      \left\{ 1 + \etax^{K}\rme^{-x_0 \Delta } + 
                  \etax^{0} \rme^{-(T-x_0) m_{\rm G} } 
      \right\} , 
      \label{e_fX_asympt} \nonumber \\[10pt]
  \FY(x_0) &\approx& 
       L^6  \, \vert \rho \vert^2 \, \langle 0,K| \opY | 0,K \rangle 
                       \, \rme^{-T m_K } \,, 
      \label{e_FY_asympt}\nonumber \\[10pt]
  f_{1} &\approx& \frac12 \rho^2 \, \rme^{-T m_K } 
                   \label{e_f1_asympt} \,, 
\ees
where we have introduced the ratios 
\bes                   
   \rho &=&  {\langle 0,K \initialK 
             \over  
             \langle 0,0 \initial } ,
                       \label{e_rho}\\[10pt]
 \etax^{K} &=& {\langle 0,0| \opX | 1, K \rangle \langle 1, K \initialK
                   \over 
                 \langle 0,0| \opX | 0, K \rangle \langle 0, K \initialK}
 \label{e_etaK}  \,, \\[10pt]
 \etax^{0} &=& { \initialt 1,0 \rangle \langle 1,0| \opX | 0, K \rangle 
                  \over
                 \initialt 0,0 \rangle \langle 0,0| \opX | 0, K \rangle }
 \label{e_eta0}  \,.
\ees
The energy difference $m_{\rm G} = E^{(0)}_{1}-E^{(0)}_{0}$ is the
mass of the $0^{++}$ glueball and $\Delta=E^{(K)}_{1}-E^{(K)}_{0}$
is an abbreviation for the gap in the $K$-meson channel.
We have dropped contributions of higher excited states which
decay even faster as $x_0$ and $T-x_0$ become large.

Considering the special case of $\fa$, one finds that it is
proportional to the matrix element $\langle 0,0| \opA |0,K \rangle$,
which is related to the kaon decay constant $\FK$ through
\bes
    \za \langle 0,0| \opA | 0,K\rangle  = 
    \FK m_K (2m_K L^3)^{-1/2} \,.
\ees
Here, $\za$ is the renormalisation constant of the isovector axial
current, and the factor $(2m_KL^3)^{-1/2}$ takes account of the
conventional normalisation of one-particle states\footnote{
We denote conventionally normalised one-Kaon states by $\vert K
\rangle$.}. In our convention
the experimental value of the pion decay constant is 
$132~\MeV$.

\Eq{e_fX_asympt} is used to determine $m_K$, while the Kaon
decay constant $\FK$ may be conveniently extracted from the ratio 
\bes
 \za \, \fa(x_0) / \sqrt{f_{1}} &\approx& 
   \frac12 \FK \, (m_K  \,L^{3})^{1/2}
             \rme^{-(x_0-T/2) m_K } \nonumber \\[10pt]
            && \times
      \left\{1 + \etaa^{K}\rme^{-x_0 \Delta } + 
                  \etaa^{0} \rme^{- (T-x_0) m_{\rm G}} 
      \right\} \, .
      \label{e_fK}
\ees
Similarly, the $\Delta S = 2$ matrix element of the operator $Y$ can
be determined from the ratio
\bes
{ {\FY (x_0) } \over {2f_{1} } } \approx
L^6 {\langle 0,K| \opY | 0,K \rangle}
= { {L^6} \over {2 m_K L^3} } {\langle K| \opY | \bar K \rangle} \,,
\nonumber
\ees
where in the last equation the factor $(2m_{K}L^3)$ again refers
to the conventional normalisation of one-particle states
$ | \bar K \rangle$ and $ | K \rangle$ in infinite volume.
Finally, the bare $B_K$ can be extracted from ratios
of the form
\bes
      \label{e_RBK}
{ {\FY (x_0) } \over {\frac{8}{3} [2\fA(x_0)] [2\fA^\prime(T-x_0)]  } }
&\approx&
{ {\langle 0,K| \opY | 0,K \rangle} \over
{ \frac{8}{3} \langle 0,0| \opA | 0,K\rangle  \langle 0,K| \opA |
  0,0\rangle} } \\[10pt]
&=& \ZA^2 \,\,
{ {\langle K| \opY | \bar K \rangle} \over {\frac{8}{3} \FK^2 m_K^2 } } \,,
\nonumber
\ees
where we define 
\bes
  \fX^\prime(T-x_0)   =  - \dfrac{L^3}{2}\langle
                 \, \op{ds}^\prime X_{ds}(x) 
         \rangle \label{e_faprime} \,,
\ees
with an analogous asymptotic expansion to that of $\fX$
(note that, for this correlation, the dominant exponential decay
is $\exp[- (T - x_0) m_K]$).
The above formulae show explicitly how masses and matrix elements
can be obtained from \SF correlation functions. A discussion on 
the practical advantages of this method of extraction of hadronic masses 
and matrix elements can be found in ref.~\cite{mbar:pap2}.

\subsection{SF correlation functions and tmQCD}
\label{sec:SFtmQCD}

Twisted mass QCD requires two modifications to the above framework.
First, the fields and the symmetries must be re-interpreted 
through the axial field transformation
as discussed in section~2. Unfortunately a direct comparison 
of SF correlation functions between tmQCD and standard QCD 
is not possible with
our current set-up, i.e. the relations~(\ref{identity_1})
and (\ref{identity_2}), connecting tmQCD and QCD renormalised correlation 
functions in an infinite volume, 
do not hold for SF correlation functions
like the ones defined in~\req{e_fX_qu} and~\req{e_fY_qu}. 
For this to work out one would need to chirally rotate both the quark mass 
terms and the quark boundary projectors at the same time 
(see\cite{Sint-tmQCD-SF} for work in this direction).  
As we do not change the fermionic boundary
projectors, the SF correlation functions for tmQCD and standard
QCD are inequivalent even in the continuum limit. 
For the quantum mechanical representation
this implies that the  initial and final states with quantum
numbers of the vacuum or a kaon state are not the same
in both cases, although we will keep the same notation.
However, the operator relations in tmQCD and standard QCD,
intended as equations between renormalised matrix elements 
of physical states, remain valid. 

Second, the exact lattice symmetries in tmQCD, which are relevant for the 
classification of excited states in a given channel,
are less restrictive, as part of the flavour symmetry group is broken, 
as well as parity. In contrast to standard Wilson quarks
one thus has to deal with excited states which have the  
wrong continuum quantum numbers, but share all the lattice
quantum numbers with the state of interest. 
A prominent example is the appearance of the
neutral pion (or kaon, depending on the definition
of the twisted doublet), which has the same lattice quantum number
as the vacuum state. Therefore, it is always possible
to have an excited state consisting of the hadron
of interest together with a zero-momentum neutral pseudoscalar meson.
On the other hand, it is clear that
the corresponding matrix elements with these states are
a pure lattice artefact suppressed by some power of the
lattice spacing. This means
that if one were to analyse the hadron spectrum 
after having taken the continuum limit of (ratios) of
correlation functions, these additional states would not
play any r\^ole. However, this procedure being somewhat impractical,
any analysis at fixed lattice spacing must deal with this problem. 

From the above discussion and Eqs.~(\ref{eq:tmA2},\ref{eq:tmVA2},\ref{eq:tmA4},\ref{eq:tmVA4},\ref{e_RBK},\ref{e_faprime})
we readily conclude that in tmQCD (with $\taa = \pi/2,\pi/4$), 
$B_K$ can be computed as the asymptotic limit of the ratio
\begin{equation}
\hat R_B = \dfrac{ i Z_{\rm VA+AV} F_{\rm VA+AV} }
{\dfrac{16}{3} [ \ZA \fA(x_0) - i \ZV \fV(x_0) ]
 [ \ZA \fA^\prime(T-x_0) - i \ZV \fV^\prime(T-x_0) ] } \,.
\label{eq:R1}
\end{equation}
The numerator of the above ratio involves the four-fermion operator
which, as explained previously, cannot be readily improved. Thus
our $B_K$ lattice estimates are affected by $\Oa$ finite cutoff
effects. For the denominator of Eq.~(\ref{eq:R1}) we use
the following expressions:
\begin{eqnarray}
\label{eq:fAimpchir1}
\ZA [\fA(x_0) + c_{\rm A} a \partial_0\fP(x_0)] - i \ZV \fV(x_0) \,, \\[10pt]
\label{eq:fAimpchir2}
\ZA [\fA^\prime(x_0) + c_{\rm A} a \partial_0\fP^\prime(x_0)] - i Z_{\rm
  V} \fV^\prime(x_0) \,.
\end{eqnarray}
Comparing the above with
Eqs.~(\ref{eq:impA2},\ref{eq:impV2},\ref{eq:impA4},\ref{eq:impV4}), we 
note that the vector current counterterm proportional to $c_{\rm T}$ vanishes
due to invariance of $f_{\rm T}$ under spatial translations.
Moreover, we have omitted the improvement counterterms proportional to 
$b_{\rm V,A}$ and $\tilde b_{\rm V,A}$.
The reason is that these factors are the same for both currents at tree level
($b_{\rm A} = b_{\rm V} = 1/2$ and $\tilde b_{\rm A} = \tilde b_{\rm V} = 0$).
It can then be easily shown that tree-level O($a$) improvement 
of the numerator $F_{\rm VA+AV}$ would require the
same terms as those of the denominator. Omitting
these counterterms from the denominator thus amounts to
a consistent cancellation of O($a m_{\rm q}$) 
effects at tree level in the ratio $\hat R_B$, while there are no
O($a\mu_{\rm q}$) effects to this order of perturbation theory. 
We have nevertheless checked that, in practical simulations, 
the contribution from the counterterms proportional to $b_{\rm A},b_{\rm V},
\tilde b_{\rm A}$ and $\tilde b_{\rm V}$ in the denominator 
is indeed negligible. 

Concerning excited states with the wrong continuum quantum 
numbers we discuss for concreteness the asymptotic behaviour of a suitable
combination of correlation functions, from which the decay
constant $F_K$ is extracted, ignoring improvement-related issues: 
\begin{eqnarray}
&& \dfrac{\sqrt 2}{2} [ \ZA \fA(x_0) - i \ZV \fV(x_0) ] \approx 
   \dfrac{L^3}{2} \rho \, \langle 0,0 \vert \opAzero \vert 0,K \rangle
             \rme^{- x_0 m_K }
 \nonumber \\[10pt]
            && \times
      \left\{1 + \etaa^{K}\rme^{-x_0 \Delta } 
               + \etaa^{0} \rme^{- (T-x_0) m_{\rm G}} 
               + \etaa^{\sigma_K}\rme^{-x_0 \Delta_\sigma }
               + \etaa^{\pi^0} \rme^{- (T-x_0) m_{\pi^0}} 
      \right\} \, .\qquad
      \label{e_fK2}
\end{eqnarray}
Here the superscript $^{(0)}$ in the axial current operator
reminds us that the renormalised operator relations corresponding 
to Eqs.~(\ref{eq:tmA2},\ref{eq:tmA4}) have been used.
As these relations become exact only in the continuum limit, besides
the $\etaa^{K}$ and $\etaa^{0}$ of Eqs.~(\ref{e_etaK}) and (\ref{e_eta0}), 
we have now included the lowest lying parity-violating contributions
\bes                   
 \etaa^{\sigma_K} &=& {\langle 0,0| \opAzero | 0, \sigma_K \rangle \langle 0, \sigma_K \initialK
                   \over 
                 \langle 0,0| \opAzero | 0, K \rangle \langle 0, K \initialK}
                   \,, \\[10pt]
 \etaa^{\pi^0} &=& { \initialt 0, \pi^0 \rangle \langle \pi^0 ,0| \opAzero | 0, K \rangle 
                  \over
                 \initialt 0,0 \rangle \langle 0,0| \opAzero | 0, K \rangle }
            \,.              
\ees
These essentially consist of:
\begin{itemize}
\item The projection of the lowest lying scalar state with a net
$s$- and $\bar d$-flavour content, denoted by $\vert  0, \sigma_K \rangle$ 
from the $x_0 = 0$ time-boundary. It has a mass $m_{\sigma_K}$ with
$\Delta_{\sigma} \equiv m_{\sigma} - m_K$.
\item The projection of the lowest lying pseudoscalar state with vacuum
(zero) net flavour content, denoted by $\vert  0, \pi^0 \rangle$ from the 
$x_0 = T$ time-boundary. It has mass $m_{\pi^0}$.\footnote{Some
care is required in the interpretation of this notation. Our $\pi/2$ tmQCD
consists of a light twisted quark doublet $\bar \psi =
(\bar u \, , \, \bar d)$ and an untwisted strange quark. So what is meant by
$\vert  0, \pi^0 \rangle$ is a state generated from the vacuum by the
operator $\bar \psi \gamma_5 \tau^3 \psi$; i.e. a pion.
Our $\pi/4$ tmQCD,
consists of a twisted quark doublet $\bar \psi = (\bar s \, , \, \bar d)$;
thus we have a theory in which physical strangeness is described by
$I_3 = \pm 1/2$ isospin quark states. We therefore have three Kaons, corresponding
to isospin values $I_3 = -1, 0, +1$. What is meant by $\vert  0, \pi^0 \rangle$
is again a state generated from the vacuum by the operator
$\bar \psi \gamma_5 \tau^3 \psi$, which is an $I_3 = 0$ Kaon.} 
\end{itemize}
In standard QCD, the relation $\etaa^{\sigma_K} = \etaa^{\pi^0} =0$ is 
rigorously satisfied on the grounds of parity conservation. As tmQCD breaks
parity, these coefficients are cutoff effects
which only vanish in the continuum limit. Based on our knowledge of QCD 
spectroscopy, we expect the scalar state $\vert  0, \sigma_K \rangle$ to
be significantly heavier than $\vert  0, K \rangle$, so it should have an
exponentially vanishing contribution. On the other hand, the state
$\vert  0, \pi^0 \rangle$, be it a true neutral pion in the $\pi/2$ theory
or an $I_3=0$ Kaon in the $\pi/4$ theory, is lighter than the glueball.
Although the associated matrix element, being $\Oa$, 
is diminishing as we approach the continuum limit, 
the exponential decay predominates on that of the 
glueball, so that the net effect may be comparable (or even dominant)
to that of the vacuum decay term. In our numerical analysis we 
obtained a rough estimate of the mass and prefactor or the first excited
state, so that  its contribution to the effective mass 
of the ground state could be estimated.

\section{Lattice $B_K$ results from tmQCD at $\taa = \pi/2$}
\label{sec:res_pi2}

The quenched $B_K$ parameter has been computed in tmQCD with a twist angle
$\taa = \pi/2$ at four values of the lattice spacing in the range
$a \approx 0.06 - 0.09 ~\fm$, with the spatial directions extending from
$L \approx 1.4$ to $L \approx 1.9 ~\fm$. Time ranges from $x_0 = 0$ to 
$x_0 = T$, with $T/L \approx 2.3-3.0$. 
The three mass parameters (i.e. the standard 
hopping parameter $\kappa_s$ for the strange quark, the twisted hopping
parameter $\kappa_d$ and the twisted mass $\mu_d$ for the down quark)
are tuned as discussed in sect.~\ref{sec:tmQCD},
so as to keep the two quarks degenerate. In order to avoid
exceptional configurations in the untwisted strange sector,
the K-meson mass is kept in the range $m_K \approx 640$ -- $830~\MeV$;
for a discussion on the presence of exceptional configurations, see
Appendix~\ref{sec:excconf}.
The parameters of our runs are displayed in Table~\ref{tab:runspi2}.
Following ref.~\cite{pot:r0}, we will express our results in units of the
length scale $r_0 \simeq 0.5 ~\fm$. For the relationship between $r_0$ and the
lattice coupling $\beta$ (which fixes the lattice spacing $a$),
see Appendix~\ref{sec:appZ}.
\begin{table}[!t]
\begin{center}
\begin{tabular}{ccccccc}
\Hline \\[-10pt]
$\beta$ & $(L/a)^3 \times T/a$ & $\frac{a}{2r_0}$ & $\frac{L}{2r_0}$  & $\kappa_s$ & 
$(\kappa_d,a\mu)$ & $N_{\rm conf}$\\
&&&&&& \\[-10pt]
\hline \\[-10pt]
6.0 & $16^3 \times 48$ & 0.0931 & 1.49 & 0.1335 & (0.135169,0.03816) & 402 \\
& & & & 0.1338 & (0.135178,0.03152) & 398 \\
& & & & 0.1340 & (0.135183,0.02708) & 402 \\
& & & & 0.1342 & (0.135187,0.02261) & 400 \\ \\[-10pt]
\hline\\[-10pt]
6.1 & $24^3 \times 56$ & 0.0789 & 1.89 & 0.1343 & (0.1354817,0.0279110) & 100 
\\
& & & & 0.1345 & (0.1354860,0.0233010) & 99 \\
& & & & 0.1347 & (0.1354896,0.0186678) & 122 \\ \\[-10pt]
\hline\\[-10pt]
6.2 & $24^3 \times 64$ & 0.0677 & 1.63 & 0.1346 & (0.1357800,0.0283240) & 200 
\\
& & & & 0.1347 & (0.1357825,0.0259850) & 201 \\
& & & & 0.1349 & (0.1357866,0.0212897) & 214 \\ \\[-10pt]
\hline\\[-10pt]
6.3 & $24^3 \times 72$ & 0.0587 & 1.41 & 0.1348 & (0.1358118,0.0246230) & 200 
\\
    &                  &       &      & 0.1349 & (0.1358139,0.0222430) & 205 
\\
    &                  &       &      & 0.1350 & (0.1358157,0.0198558) & 175 
\\
    &                  &       &      & 0.1351 & (0.1358174,0.0174640) & 201 
\\ \\[-10pt]
\Hline
\end{tabular}
\end{center}
\caption{The parameters of the runs at twist angle $\taa = \pi/2$.}
\label{tab:runspi2}
\end{table}

The effective pseudoscalar meson mass is computed from the ratio
\begin{equation}
a M^{\rm eff}_{\rm PS} (x_0) = \frac{1}{2}
\ln \Bigg [ \dfrac{ f_{\rm A_R} (x_0-a) -i f_{\rm V_R} (x_0-a)}
{f_{\rm A_R} (x_0+a) -i f_{\rm V_R} (x_0+a)} \Bigg ] \,,
\label{eq:effmass}
\end{equation}
where $f_{\rm A_R}$, $f_{\rm V_R}$ are the correlation functions constructed
from the improved operators of Eqs.~(\ref{eq:impA2}) and (\ref{eq:impV2}),
giving results which are free of $\Oa$ effects at all times.
In order to increase the signal stability, these correlation functions
are antisymmetrised with their partners $f^\prime_{\rm A_R}$ and
$f^\prime_{\rm V_R}$.

Flavour breaking effects have also been monitored by comparing the
effective pseudoscalar meson mass of two axial correlation functions.
The first is the quantity defined above, derived from the correlation function
composed of a strange (standard Wilson) and a down (tmQCD Wilson)
quark propagators. The second is the effective mass derived from
the fully twisted correlator,
composed of an up and a down (tmQCD Wilson) quark propagator.
As the twisted and untwisted quark flavours have been tuned
to be degenerate in mass to $O(a^2)$
(cf. subsect.~\ref{sec:masstune}), this is a measure of flavour
breaking effects.
The two effective masses computed from these correlation functions
differ by $\sim 5\%$ at $\beta = 6.0$, while at $\beta = 6.3$
they  practically coincide. Thus, small flavour breaking effects,
visible at large lattice cutoffs, disappear as we move towards
the continuum limit. Nevertheless, these effects require further and
more detailed investigation.

In order to determine the plateaux of the effective masses we have
followed the procedure of ref.~\cite{mbar:pap2}. We have allowed for
a relative excited state contribution to the effective mass of at most 
$0.2\%$ within the plateaux. The plateaux of the effective masses
$[x_0^{\rm min}/r_0, x_0^{\rm max}/r_0]$, which satisfy this criterion,
are listed in Table~\ref{tab:respi2} and illustrated in
Fig.~\ref{fig:plateaux_m_pi2_1}. The values for
$x_0^{\rm min}/r_0$ indicate that the K-meson decay channel dominates
the excited states after roughly $1.25 - 1.35 ~\fm$ from the $x_0 = 0$
wall source. The pseudoscalar meson mass $a M_{\rm PS}$ is obtained
by averaging the $a M^{\rm eff}_{\rm PS}(x_0)$ values in the plateaux;
errors are estimated by the jackknife procedure, omitting one measurement
from each bin. We present this result in the form $r_0 M_{\rm PS}$
in Table~\ref{tab:respi2}.

Given these considerations, we determine $B_K$ by averaging in the symmetric interval
$[x_0^{\rm min}/r_0, (T-x_0^{\rm min})/r_0]$. What we compute is the quantity
$R_B \equiv \hat R_B/Z_{\rm VA+AV}$ (cf. Eqs.~(\ref{eq:R1},\ref{eq:fAimpchir1},
\ref{eq:fAimpchir2})), corresponding to the bare $B_K$.\footnote{
Recall that the numerator of $R_B$ is the bare correlation function
$F_{VA+AV}$ while the denominator is a properly normalised combination
of $f_A$ and $f_V$. Also recall that the ratio is improved at tree level, 
and the denominator is $\Oa$ improved in the chiral limit.}
We distinguish three sources of uncertainty in our $R_B$ results:
\begin{enumerate}
\item The results depend on the prescription used for the
determination of the normalisation constants $\ZA$ and $\ZV$, as well
as the improvement coefficient $\icA$. As discussed in
Appendix~\ref{sec:appZ}, two such independent
determinations have been provided by the
ALPHA ~\cite{impr:pap3,impr:pap4} and LANL~\cite{lanl:imppap,lat01:gupta}
collaborations.\footnote{The values provided by the LANL collaboration in the more recent ref.~\cite{Bhattacharya:2005ss} only change within the quoted errors with respect to the ones we have used in our analysis.} We compute the ratio $R_B$ for both sets of values. The
discrepancy between the two determinations is a measure of $O(a^2)$
systematic effects. We also compute $R_B$ with the ALPHA collaboration
estimates for $\ZA$ and $\ZV$, but without the improving $\icA$
counterterm in the axial current. Any discrepancy between this and the
previous determinations of $R_B$ signals the presence of strong $\Oa$
effects.
\item The statistical errors of the correlation functions are computed
by a standard jackknife error analysis (omitting one measurement from
each bin), without taking into account the errors of $\ZA$, $\icA$ and $\ZV$.
\item The total statistical error is computed by combining the statistical
errors of $\ZA$, $\ZV$ (for a given ALPHA or LANL determination)
with those of the correlation functions, through the standard error
propagation procedure. The statistical error of $\icA$ is not taken into 
account, as it is related to an $\Oa$ correction (recall that for the same
reason, $\icsw$ is also used without an error).
\end{enumerate}

\begin{table}[!t]
\begin{center}\scriptsize
\begin{tabular}{cccccc}
\Hline \\[-6pt] 
$\beta$ & $\big [ \dfrac{x_0^{\rm min}}{2r_0}, \dfrac{x_0^{\rm max}}{2r_0} \big ]$ &
$r_0 M_{\rm PS}$ & $R_B^{\rm ALPHA}$ & $R_B^{\rm LANL}$  & $R_B^{{\rm
  ALPHA;~w/o}~\icA}$ \\
&&&&& \\[-6pt]
\hline \hline \\[-6pt]
6.0 & $[1.30,3.17]$ & 2.092(6) & 1.177(9)(8)(12) & 1.089(8)(11)(14) & 1.025(8)(6)(10) \\ \\[-6pt]
    & & 1.907(7) & 1.139(13)(8)(15) & 1.054(12)(11)(16) & 0.993(12)(6)(13) \\ \\[-6pt]
    & & 1.780(6) & 1.123(14)(8)(16) & 1.038(12)(11)(16) & 0.977(12)(6)(13) \\ \\[-6pt]
    & & 1.635(6) & 1.084(13)(8)(15) & 1.001(12)(10)(16) &
0.943(11)(5)(12) \\ \\[-6pt]
\hline \\[-6pt]
    & & 1.2544 & 1.031(19)(23) & 0.952(17)(24) & 0.896(16)(18) \\ \\[-6pt]
\hline \hline \\[-6pt]
6.1 & $[1.34,3.08]$ & 1.978(6) & 1.066(9)(8)(12) & 1.041(8)(7)(11) & 0.992(8)(6)(10) \\ \\[-6pt]
    & & 1.812(7) & 1.023(10)(7)(12) & 0.998(10)(6)(12) & 0.951(9)(6)(11) \\ \\[-6pt]
    & & 1.647(6) & 0.981(10)(7)(12) & 0.955(10)(6)(12) &
0.911(9)(5)(10) \\ \\[-6pt]
\hline \\[-6pt]
    & & 1.2544 & 0.901(21)(25) & 0.874(20)(25) & 0.835(18)(21) \\ \\[-6pt]
\hline \hline \\[-6pt]
6.2 & $[1.29,3.05]$ & 2.079(6) & 1.023(7)(7)(10) & 1.017(7)(8)(11) & 0.980(7)(6)(9) \\ \\[-6pt]
    & & 1.980(7) & 0.993(9)(7)(11) & 0.987(9)(7)(11) & 0.951(8)(5)(9) \\ \\[-6pt]
    & & 1.795(7) & 0.975(8)(7)(11) & 0.970(8)(7)(11) &
0.935(8)(5)(9) \\ \\[-6pt]
\hline \\[-6pt]
    & & 1.2544 & 0.902(23)(31) & 0.900(23)(32) & 0.868(23)(26) \\ \\[-6pt]
\hline \hline \\[-6pt]
6.3 & $[1.23,3.00]$ & 2.050(6) & 0.990(12)(7)(14) & 0.996(12)(5)(13) & 0.962(12)(5)(13) \\ \\[-6pt]
    & & 1.962(8) & 0.996(15)(7)(17) & 1.003(15)(5)(16) & 0.967(14)(5)(15) \\ \\[-6pt]
    & & 1.839(10) & 0.959(17)(7)(18) & 0.964(17)(5)(18) &0.932(16)(5)(17) \\ \\[-6pt]
    & & 1.722(9) & 0.918(15)(6)(16) & 0.924(15)(5)(16) &
 0.892(14)(5)(15) \\ \\[-6pt]
\hline \\[-6pt]
    & & 1.2544 & 0.844(32)(35) & 0.851(32)(34) & 0.820(30)(33) \\ \\[-6pt]
\Hline \\[-6pt]
\end{tabular}
\end{center}
\caption{
Results for the pseudoscalar mass and the various ratios from which
$B_K$ is extracted (twist angle $\taa = \pi/2$). The error in $r_0 M_{\rm PS}$ is statistical. The three errors
of the $R_B$ ratio are, in order of appearance: (i) due to the statistical
fluctuations of the correlations; (ii) due to the errors of 
$Z_{\rm A,V}$; (iii) the total error from the two previous ones. The results of the
extrapolations to the physical Kaon mass values are shown at the bottom of
each $\beta$-dataset: the first error of $R_B$ is that arising from
type-(i) errors of the fitted values, while the second is from type-(iii) errors.}
\vskip 0.2cm
\label{tab:respi2}
\end{table}

The results for $R_B$ (and pseudoscalar masses) are collected in
Table~\ref{tab:respi2}. They have all been computed at the
time-plateaux shown in the second column, determined
from the effective mass
of \req{eq:effmass} with ALPHA collaboration values for $\ZA$, $\ZV$
and $\icA$. The quality of the data is also illustrated in
\refig{fig:plateaux_m_pi2_1}. It is clear that as the continuum
limit is approached, the $O(a^2)$ discrepancies between $R_B^{\rm ALPHA}$
and $R_B^{\rm LANL}$ tend to decrease. This tendency is less marked
for $R_B^{{\rm ALPHA~w/o}~\icA}$, which differs from the other two by 
$\Oa$ effects.

The $R_B$ values are extrapolated linearly in $(r_0 M_{\rm PS})^2$, to
the physical point $(r_0 M_K)^2 = 1.5736$
(corresponding to $M_K^2 = \half[M_{K^0}^2+M_{K^\pm}^2]$), see \refig{fig:extrap_pi2}.
An interesting issue concerns the magnitude of the error of these extrapolated
values. Usually, in quenched simulations performed at a given $\beta$,
observables such as the ratio $R_B$ are computed on the same
configuration ensemble for all values of the bare quark masses (hopping parameters). This
means that the extrapolation to the strange quark mass is performed on
data points which are strongly correlated, thus reducing the error
of the extrapolation. We have instead opted for independent Monte Carlo
simulations for each set of bare quark masses, which in this respect mimic 
unquenched simulations, at the price of having uncorrelated $R_B$ results
with larger errors on the extrapolations.

The justification of the linear fit of  $R_B$, as a function of the
squared pseudoscalar mass is provided in Appendix~\ref{sec:applinfit}.

\subsection{Finite volume effects}

In order to investigate finite volume effects, we have performed a simulation
at $\beta = 6.0$ for a larger spatial volume, with extension $L/a = 24$,
at the lightest mass $r_0 M_{\rm PS} = 1.635$.
The results are gathered in Table~\ref{tab:vol2}.
\begin{table}[!t]
\begin{center}
\begin{tabular}{cccc}
\Hline \\[-10pt]
$L^3 \times T /a^4$ & $r_0 M_{\rm PS}$ & $R_B^{\rm ALPHA}$ &
$N_{\rm conf}$ \\ \\[-10pt]
&&& \\[-10pt]
\hline \\[-10pt]
$16^3 \times 48$ & 1.635(6) & 1.084(13)(8)(15) & 400 \\
$24^3 \times 48$ & 1.624(5) & 1.079(7)(8)(11) & 167 \\ \\[-10pt]
\Hline
\end{tabular}
\end{center}
\caption{$B_K$-parameter at fixed effective pseudoscalar mass $M_{\rm PS}^{\rm eff}$
for two different spatial lattice volumes (twist angle $\taa = \pi/2$). The various errors on $R_B$
are explained in the caption of Table~\ref{tab:respi2}.}
\label{tab:vol2}
\end{table}
Finite volume effects are estimated in terms of the following ratios:
\begin{eqnarray}
\dfrac{M_{\rm PS}(L/a = 16)}{M_{\rm PS}(L/a = 24)} 
-1 = 0.007(5) \,, \qquad
\dfrac{R_B^{\rm ALPHA}(L/a = 16)}{R_B^{\rm ALPHA}(L/a = 24)} -1 = 0.005(17) \,.
\end{eqnarray}
Note that all errors are of purely statistical nature.
We see that the effective mass ratios have a very small statistical deviation 
from zero, while the $B_K$ ratio is essentially insensitive to finite volume 
effects.

\subsection{Non-degenerate masses}

Past and current simulations for the determination of $B_K$ have mostly
been carried out with degenerate down and strange quarks. The rationale
behind this (at least as far as quenched simulations are concerned) is
to be found in the chiral perturbation theory expression for $B_K$. As
discussed in ref.~\cite{sharpe:Qchlogs,rev:sharpe,SharpeZhang,GoltLeung1}, 
the quenched chiral expression for $B_K$ is given by
\begin{equation}
B_K = B \Big [ 1 - (3 + \epsilon^2) y \ln y + b y + c y  \epsilon^2
+ \delta \Big \{ \dfrac{2 - \epsilon^2}{2 \epsilon} 
\ln \left( \dfrac{1 - \epsilon}{1 + \epsilon} \right) + 2 \Big \} \Big ] \, ,
\label{eq:BKchPT}
\end{equation}
with $y = m_K^2/(4\pi F)^2$ and
$F \simeq 102$\,MeV (see ref.~\cite{Kpipi:coupl2}). The parameter
\begin{equation}
\epsilon \equiv \dfrac{m_s - m_d}{m_s + m_d}
\end{equation}
is a measure of the down-strange degeneracy breaking. 
The expression~(\ref{eq:BKchPT}) is identical in form to the
one valid for dynamical quarks, save for the $\delta-$term.
This quenched artefact 
vanishes in the degenerate limit $\epsilon \rightarrow 0$, but diverges
when the down quark becomes chiral and $\epsilon \rightarrow 1$.

We have measured $R_B$ with non-degenerate down and strange
quarks. Our aim is to probe its dependence on quark mass differences,
while avoiding the potentially dangerous $\epsilon \rightarrow 1$ limit.
We have performed runs at $\beta = 6.0$, for two values of $\epsilon \ne 0$,
tuning the bare mass parameters so that the pseudoscalar mass remains close
to the value $r_0M_{\rm PS} = 1.780$ of the $\epsilon = 0$ simulation.
Our results are summarised in Table~\ref{tab:eps}.
\begin{table}[!t]
\begin{center}
\begin{tabular}{cccccc}
\Hline \\[-10pt] 
$\epsilon$ & $\kappa_s$ & ($\kappa_d$,$a \mu$) & $r_0 M_{\rm PS}$
& $R_B^{\rm ALPHA}$ & $N_{\rm conf}$
\\ \\[-10pt]
\hline \\[-10pt]
0.00 & 0.1340 & (0.1351830,0.02708) & 1.780(6) & 1.123(14)(8)(16) & 400 \\
0.16 & 0.1338 & (0.1351867,0.02259) & 1.775(6) & 1.128(17)(8)(19) & 400 \\
0.41 & 0.1335 & (0.1351913,0.01598) & 1.772(7) & 1.129(14)(8)(16) & 402
\\ \\[-10pt]
\Hline
\end{tabular}
\end{center}
\caption{$B_K$-parameter at fixed effective pseudoscalar mass $M_{\rm PS}^{\rm eff}$
for three values of the degeneracy-breaking parameter $\epsilon$ (twist angle $\taa = \pi/2$). The various errors on $R_B$
are explained in the caption of Table~\ref{tab:respi2}.}
\label{tab:eps}
\end{table}
The effect of breaking the mass degeneracy of the valence quarks is estimated
in terms of the following ratios:
\begin{align}
\dfrac{M_{\rm PS}(\epsilon = 0)}{M_{\rm PS}(\epsilon = 0.16)} 
-1 & = 0.003(5),  & \dfrac{M_{\rm PS}(\epsilon = 0)}{M_{\rm PS}(\epsilon=0.41)}  -1 & = 0.005(5),
\\[10pt]
\dfrac{\ \ R_B^{\rm ALPHA}(\epsilon = 0)\ \ }{R_B^{\rm ALPHA}(\epsilon = 0.16)} -1 & = -0.004(19),  &
\dfrac{\ \ R_B^{\rm ALPHA}(\epsilon = 0)\ \ }{R_B^{\rm ALPHA}(\epsilon = 0.41)} -1 & = -0.005(17).
\end{align}
All errors are statistical.
We see that there is no appreciable deviation between $\epsilon = 0$ and 
$\epsilon \ne 0$ values. It seems that, at least in the region of mass 
differences explored, $B_K$ is not sensitive to the breaking of mass 
degeneracy.

Finally, we find it worth to mention in this context a curious by-product of 
our computations that concerns the somewhat surprising appearance of an 
exceptional configuration in the $\pi/2$ simulation with non-twisted strange 
valence quarks, corresponding to a rather heavy $K$-meson of 
$m_K \simeq 650$ MeV.
This happened at $\beta = 6.0$ and for a ``standard'' value of the Wilson
hopping parameter which, being generally considered ``safe'' from
exceptional configurations, has also been used in the simulations of
other collaborations.
More details are provided in Appendix~\ref{sec:excconf}.

\section{Lattice $B_K$ results from tmQCD at $\taa = \pi/4$}
\label{sec:res_pi4}

\begin{table}
[!t]
\begin{center}
\begin{tabular}{cccccc}
\Hline \\[-10pt] 
$\beta$ & $(L/a)^3 \times T/a$ & $\frac{a}{2r_0}$ & $\frac{L}{2r_0}$ & $(\kappa,a\mu)$ & $N_{\rm conf}$\\ 
\\[-10pt]
\hline \\[-10pt]
6.00 & $24^3 \times 48$ & 0.0931 & 2.24 & (0.134739,0.010412) & 200 \\
& & & & (0.134795,0.009142) & \\
& & & & (0.134828,0.008397) & \\ \\[-10pt]
\hline \\[-10pt]
6.10 & $24^3 \times 60$ & 0.0789 & 1.89 & (0.135152,0.00810) & 200 \\
& & & & (0.135190,0.00720) & \\
& & & & (0.135235,0.00615) & \\ \\[-10pt]
\hline \\[-10pt]
6.20 & $32^3 \times 72$ & 0.0677 & 2.17 & (0.135477,0.007595) & 73 \\
& & & & (0.135539,0.006125) & \\ \\[-10pt]
\hline \\[-10pt]
6.30 & $32^3 \times 72$ & 0.0587 & 1.88 & (0.135509,0.0076) & 76 \\
& & & & (0.135546,0.0067) & \\
& & & & (0.135584,0.0058) & \\ \\[-10pt]
\hline \\[-10pt]
6.45 & $32^3 \times 86$ & 0.0481 & 1.54 & (0.135105,0.01459) & 105 \\
& & & & (0.135218,0.01185) & \\
& & & & (0.135293,0.01002) & \\ \\[-10pt]
\Hline
\end{tabular}
\end{center}
\caption{The parameters of the runs at twist angle $\taa = \pi/4$.}
\label{tab:runspi4}
\end{table}

\begin{table}[!t]
\begin{center}\scriptsize
\begin{tabular}{cccccccc}
\Hline \\[-6pt] 
$\beta$ & $\big [ \dfrac{x_0^{\rm min}}{2r_0}, \dfrac{x_0^{\rm max}}{2r_0} 
\big ]$ &
$r_0 M_{\rm PS}$ & $R_B^{\rm ALPHA}$ & $R_B^{\rm LANL}$  & $R_B^{{\rm
  ALPHA;~w/o}~\icA}$ \\ \\[-6pt]
\hline \\[-6pt]
6.0 & $[1.30,3.17]$ & 1.326(4) & 1.067(8)(8)(11) & 0.960(7)(10)(12) & 0.885(6)
(5)(8) \\ \\[-6pt]
    & $[1.30,3.17]$ & 1.253(4) & 1.048(8)(7)(11) & 0.941(7)(9)(11) & 0.868(7)
(5)(9) \\ \\[-6pt]
    & $[1.30,3.17]$ & 1.207(4) & 1.036(9)(7)(11) & 0.930(8)(9)(12) & 0.857(7)
(5)(9) \\ \\[-6pt]
\hline \\[-6pt]
    & & 1.2544 & 1.048(8)(11) & 0.943(7)(12) & 0.869(7)(9) \\ \\[-6pt]
\hline\hline \\[-6pt]
6.1 & $[1.18,3.55]$ & 1.381(5) & 0.912(8)(7)(11)  & 0.883(8)(6)(10) & 0.829
(8)(5)(9) \\ \\[-6pt]
    & $[1.18,3.55]$ & 1.325(5) & 0.892(9)(6)(11)  & 0.863(9)(6)(11) & 0.810
(8)(5)(9) \\ \\[-6pt]
    & $[1.26,3.47]$ & 1.257(5) & 0.865(10)(6)(12) & 0.836(9)(6)(11) & 0.785
(9)(5)(10) \\ \\[-6pt]
\hline \\[-6pt]
    & & 1.2544 & 0.865(10)(11) & 0.836(9)(11) & 0.784(9)(10) \\ \\[-6pt]
\hline\hline \\[-6pt]
6.2 & $[1.35,3.52]$ & 1.299(6) & 0.882(11)(6)(13) & 0.875(11)(6)(13) & 0.831
(10)(5)(11) \\ \\[-6pt]
    & $[1.35,3.52]$ & 1.182(6) & 0.852(12)(6)(13) & 0.845(12)(6)(13) & 0.803
(11)(5)(12) \\ \\[-6pt]
\hline \\[-6pt]
    & & 1.2544 & 0.869(11)(13) & 0.863(10)(12) & 0.818(10)(11) \\ \\[-6pt]
\hline\hline \\[-6pt]
6.3 & $[1.29,2.94]$ & 1.338(9)  & 0.848(13)(6)(14) & 0.854(13)(5)(14) & 0.815
(12)(5)(13) \\ \\[-6pt]
    & $[1.35,2.88]$ & 1.259(9)  & 0.831(14)(6)(15) & 0.835(14)(5)(15) & 0.798
(13)(5)(14) \\ \\[-6pt]
    & $[1.47,2.76]$ & 1.175(10) & 0.811(15)(6)(16) & 0.814(14)(5)(15) & 0.779
(14)(4)(15) \\ \\[-6pt]
\hline \\[-6pt]
    & & 1.2544 & 0.829(13)(14) & 0.834(13)(14) & 0.797(12)(13) \\ \\[-6pt]
\hline\hline \\[-6pt]
6.45& $[1.30,2.84]$ & 2.054(10) & 0.937(11)(7)(13) & 0.949(10)(6)(13) & 0.913
(10)(5)(11) \\ \\[-6pt]
    & $[1.25,2.88]$ & 1.848(11) & 0.904(12)(6)(13) & 0.915(12)(6)(13)  & 0.881
(11)(5)(12) \\ \\[-6pt]
    & $[1.30,2.84]$ & 1.702(11) & 0.879(13)(6)(14) & 0.889(13)(6)(14)  & 0.856
(13)(5)(14) \\ \\[-6pt]
\hline \\[-6pt]
    & & 1.2544 & 0.821(16)(18) & 0.831(16)(17) & 0.800(16)(17) \\ \\[-6pt]
\Hline
\end{tabular}
\end{center}
\caption{Results for the pseudoscalar mass and the various ratios from which
$B_K$ is extracted (twist angle $\taa = \pi/4$). The error in $r_0 M_{\rm PS}$ is statistical. The three 
errors of the $R_B$ ratio are, in order of appearance: 
(i) due to the statistical
fluctuations of the correlations; (ii) due to the errors of 
$Z_{\rm A,V}$; (iii) the total error from the two previous ones. The results of 
the
extrapolations to the physical Kaon mass values are shown at the bottom of
each $\beta$-dataset: the first error of $R_B$ is that arising from
type-(i) errors of the fitted values, while the second is from type-(iii) 
errors.}
\label{tab:respi4}
\end{table}

The quenched $B_K$ parameter has also been computed in tmQCD with a twist angle
$\taa = \pi/4$. This formalism, with twisted down and strange quarks in
the same flavour doublet, is a way to avoid the problem of exceptional
configurations, while maintaining the property of multiplicative
renormalisation of the four-fermion operator. Therefore, contrary to the
$\taa = \pi/2$ simulation, in the present one it is possible to simulate
at half the value of the physical strange quark mass directly, thus avoiding 
the uncertainties introduced by extrapolations from higher masses. 
The price to pay is that larger physical
volumes are required at these smaller masses. Our simulations have been performed at four values of the lattice spacing in the range
$a \approx 0.05 - 0.09 ~\fm$, while lattice spatial extensions are
$L \approx 1.9 - 2.2 ~\fm$. Time ranges from $x_0 = 0$ to $x_0 = T$,
with $T/L \ge 2$. The two mass parameters (hopping parameter $\kappa$ 
and twisted mass $\mu$ for the two degenerate valence flavours)
are tuned as discussed in sect.~\ref{sec:tmQCD}. 
Contrary to the $\pi/2$ case, here we have generated a single configuration
ensemble per $\beta$, on which measurements at a few $(\kappa, a \mu)$
values, tuned to be close to half the strange quark mass, are performed.
The parameters of our runs are displayed in Table~\ref{tab:runspi4}.
We point out that at the smallest lattice spacing, corresponding to
$\beta = 6.45$,  APEmille memory limitations 
made it impossible to maintain the physical
volume used at the lower $\beta$ values. Consequently, in this case
we were forced to run at a length of $L \approx 1.5 ~\fm$, with larger masses,
and extrapolate to the pseudoscalar mass at the physical kaon mass.

The effective pseudoscalar meson mass is computed from the ratio of
Eq.~(\ref{eq:effmass}); again correlation functions are antisymmetrised
in time. The excited states analysis, carried out as in
ref.~\cite{mbar:charm1}, determines the effective mass plateaux
for which the relative contribution from the excited states is at most
$0.5\%$.
The plateaux of the effective masses $[x_0^{\rm min}/r_0, x_0^{\rm max}/r_0]$
are listed in Table~\ref{tab:respi4} and illustrated in
Fig.~\ref{fig:plateaux_m_pi4_1}. Once more, they have been
obtained with the ALPHA collaboration values for $\ZA$, $\ZV$ and $\icA$.
The quoted values for
$x_0^{\rm min}/r_0$ indicate that the K-meson channel dominates
the excited states after roughly  $1.2$ to $1.4~\fm$ from the $x_0 = 0$
wall source. The pseudoscalar meson mass $a M_{\rm PS}$ is obtained
by averaging the $a M^{\rm eff}_{\rm PS}(x_0)$ values in the plateaux;
errors are estimated by the jackknife procedure, omitting one measurement
from each bin. This result is presented in the form $r_0 M_{\rm PS}$
in Table~\ref{tab:respi4}.

The ratio $R_B$ is computed in the symmetric interval
$[x_0^{\rm min}/r_0, (T-x_0^{\rm min})/r_0]$.
The results for $R_B$ are collected in Table~\ref{tab:respi4}.
We see once more that as the continuum limit is approached,
the $O(a^2)$ discrepancies between $R_B^{\rm ALPHA}$
and $R_B^{\rm LANL}$ tend to decrease. This tendency is less marked
for $R_B^{{\rm ALPHA~w/o}~\icA}$, which differs from the other two by 
$\Oa$ effects.
The plateaux from which $B_K$ is extracted are illustrated
in Fig.~\ref{fig:plateaux_BK_Alpha_pi4}.

The $R_B$ values are interpolated linearly in $(r_0 M_{\rm PS})^2$,
to the physical point $(r_0 M_K)^2 = 1.5736$, see \refig{fig:interp_pi4}.
These interpolations are very short and involve data which, having
been obtained at the same configuration ensemble, are correlated.
Thus, the error on the interpolated value, compared to the one obtained
in the $\taa = \pi/2$ case by extrapolation from higher pseudoscalar masses,
is smaller. For the $\beta = 6.45$ case, we still have correlated data,
but as the physical value is reached by an extrapolation, the error
is relatively large.

\subsection{Finite volume effects}

In order to test finite volume effects, we have performed simulations
at $\beta = 6.0, 6.1, 6.2$ for various spatial volumes and light quark masses.
The simulation points, as well as the results for the
pseudoscalar meson mass and for $R_B$, are gathered in
Table~\ref{tab:vol4}. The spatial extent of the lattices considered
roughly ranges from $1.5$ to $3.0$~fm.
\begin{table}[!t]
\begin{center}\scriptsize
\begin{tabular}{cccccccc}
\Hline \\[-6pt]
$\beta$ & $L^3 \times T /a^4$ & $\dfrac{L}{2 r_0}$ & $\dfrac{T}{2 r_0}$ 
& $(\kappa, a \mu)$ & $r_0 M_{\rm PS}$ & $R_B^{\rm ALPHA}$ & $N_{\rm conf}$ \\ \\[-6pt]
\hline \\[-6pt]
6.0 & $16^3 \times 48$ & 1.488 & 4.464 & $(0.134739, 0.010412)$ & 1.316(5) & 1.040(11)(7)(13) & 667 \\ \\[-6pt]
    & $16^3 \times 48$ & 1.488 & 4.464 & $(0.134828, 0.008397)$ & 1.196(6) & 0.999(13)(7)(15) & 667 \\ \\[-6pt]
    & $24^3 \times 48$ & 2.232 & 4.464 & $(0.134739, 0.010412)$ & 1.326(4) & 1.067(8)(8)(11)  & 200 \\ \\[-6pt]
    & $24^3 \times 48$ & 2.232 & 4.464 & $(0.134828, 0.008397)$ & 1.207(4) & 1.036(9)(7)(11)  & 200 \\ \\[-6pt]
    & $32^3 \times 60$ & 2.976 & 5.580 & $(0.134828, 0.008397)$ & 1.201(3) & 1.035(6)(7)(9)   & 76 \\  \\[-6pt]
\\[-6pt] \hline \\[-6pt]
6.1 & $24^3 \times 60$ & 1.896 & 4.740 & $(0.135152, 0.008100)$ & 1.381(5) & 0.912(8)(7)(11) & 200 \\  \\[-6pt]
    & $24^3 \times 60$ & 1.896 & 4.740 & $(0.135235, 0.006150)$ & 1.257(5) & 0.865(9)(6)(11) & 200 \\  \\[-6pt]
    & $32^3 \times 64$ & 2.528 & 5.056 & $(0.135152, 0.008100)$ & 1.370(5) & 0.925(7)(7)(10) & 77 \\   \\[-6pt]
    & $32^3 \times 64$ & 2.528 & 5.056 & $(0.135235, 0.006150)$ & 1.245(5) & 0.876(8)(6)(10) & 77 \\   \\[-6pt]
\hline \\[-6pt]
6.2 & $24^3 \times 72$ & 1.620 & 4.860 & $(0.135477, 0.007595)$ & 1.298(8) & 0.853(16)(6)(17) & 191 \\ \\[-6pt]
    & $24^3 \times 72$ & 1.620 & 4.860 & $(0.135539, 0.006125)$ & 1.182(9) & 0.813(18)(6)(19) & 191 \\ \\[-6pt]
    & $32^3 \times 72$ & 2.160 & 4.860 & $(0.135477, 0.007595)$ & 1.299(6) & 0.882(11)(6)(13) & 73 \\  \\[-6pt]
    & $32^3 \times 72$ & 2.160 & 4.860 & $(0.135539, 0.006125)$ & 1.182(6) & 0.852(12)(6)(13) & 73 \\  \\[-6pt]
\Hline
\end{tabular}
\end{center}
\caption{$B_K$ parameter (at two values of the effective pseudoscalar
  mass $M_{\rm PS}$ per $\beta$)
  for different spatial lattice volumes (twist angle $\taa = \pi/4$). The various errors on $R_B$
  are explained in the caption of Table~\ref{tab:respi4}.}
\label{tab:vol4}
\end{table}
The usual $M_{\rm PS}$ and $R_B^{\rm ALPHA}$ ratios at two different
spatial volumes are computed at fixed $\beta$ and their deviation from
unity is presented in Table~\ref{tab:vol4b}.

For $\beta = 6.0, 6.1$ the effective mass ratios display some small
finite volume effects, at $L \sim 1.5$~fm to $2$~fm, which appear to
vanish at $L \sim 2.5$~fm. For $\beta = 6.2$ however, such effects are
also absent already at $L \sim 1.5$~fm.
It has to be noted, however, that at this level of precision
it is difficult to disentangle finite volume effects from e.g.
the difference in cutoff effects, or
the systematic uncertainties related to excited states.

For $B_K$ we see finite volume effects at
$L \sim 1.5$~fm, which disappear at $L \sim 2$~fm. This situation
is $\beta$-independent.

\begin{table}[!t]
\begin{center}\scriptsize
\begin{tabular}{cccccccc}
\Hline \\[-6pt]
$\beta$ & $\dfrac{L_1}{a}$ & $\dfrac{L_2}{a}$ & $\dfrac{L_1}{2 r_0}$ & $\dfrac{L_2}{2 r_0}$ 
& $(\kappa, a \mu)$ &
$\left[\dfrac{M_{\rm PS}(L_1)}{M_{\rm PS}(L_2)}-1\right]$
& 
$\left[\dfrac{R_B^{\rm ALPHA}(L_1)}{R_B^{\rm ALPHA}(L_2)}-1\right]$
\\ \\[-6pt]
\hline \\[-6pt]
6.0 & 16 & 24 & 1.488 & 2.232 & $(0.134739, 0.010412)$ & -0.008(5) & -0.023(16) \\ \\[-6pt]
    &    &    &       &       & $(0.134828, 0.008397)$ & -0.009(6) & -0.033(18) \\ \\[-6pt]
    & 24 & 32 & 2.232 & 2.976 & $(0.134828, 0.008397)$ &  0.005(4) & -0.002(14) \\ \\[-6pt]
\\[-6pt] \hline \\[-6pt]
6.1 & 24 & 32 & 1.896 & 2.528 & $(0.135152, 0.008100)$ & 0.008(5) & -0.014(16) \\  \\[-6pt]
    &    &    &       &       & $(0.135235, 0.006150)$ & 0.010(6) &  0.014(17) \\  \\[-6pt]
\hline \\[-6pt]
6.2 & 24 & 32 & 1.620 & 2.160 & $(0.135477, 0.007595)$ & -0.001(8) & -0.032(23) \\ \\[-6pt]
    &    &    &       &       & $(0.135539, 0.006125)$ & 0.000(10) & -0.046(26) \\ \\[-6pt]
\Hline
\end{tabular}
\end{center}
\caption{Finite volume effects for pseudoscalar masses $M_{\rm PS}$
  and $B_K$ parameters (twist angle $\taa = \pi/4$). The errors are statistical.}
\label{tab:vol4b}
\end{table}
 
\section{$B_K$ in the continuum limit}
\label{sec:bkcon}

Having obtained the value of the bare $B_K$ parameter at several lattice 
spacings, we now proceed to the computation of its renormalised counterpart 
in the continuum limit. In ref.~\cite{ssf:vapav}, the necessary 
renormalisation constants have been calculated non-perturbatively
in various Schr\"odinger functional schemes, as well as the operator step
scaling functions, which determine its renormalisation group running.
What we need from ref.~\cite{ssf:vapav} is the total renormalisation
factor $\ZtotVApAV{;}^+ (g_0)$, which connects the bare $B_K$ to its
RGI value as follows:
\begin{equation}
\hat B_K = 
\lim_{g_0 \rightarrow 0} \,\,\, \ZtotVApAV{;s}^+ (g_0) \,\,\,B_K(g_0)\,.
\label{eq:BKcont}
\end{equation}
The bare $B_K(g_0)$ is given by $R_B$ at 
the physical point $r_0 M_K$ (see Tables~\ref{tab:respi2}
and~\ref{tab:respi4}). Which of the three candidates $R_B^{\rm ALPHA}$,
$R_B^{\rm LANL}$ or $R_B^{{\rm ALPHA;~w/o}~\icA}$ is most suitable will
be discussed below.
The quantity $\ZtotVApAV{;s}^+$ is the product of three factors:
(i) the operator renormalisation constant $\ZVApAV{;s}^+(g_0, a\mumin)$ 
at a hadronic scale $\mumin$;
(ii) the RG-running evolution function from the scale $\mumin$ to a (hopefully)
perturbative scale $\mumax = 2^7 \mumin$; (iii) the RG evolution function
from the scale $\mumax$ to infinity. For several SF renormalisation schemes,
the first two factors are known
non-perturbatively (see ref.~\cite{ssf:vapav}), while the third is
known in NLO perturbation theory (see ref.~\cite{ssf:vapavPT}).
Following ref.~\cite{ssf:vapav} we write
\begin{equation}
\ZtotVApAV{;s}^+ (g_0) = 
\zrgiVApAV{;s}^+(\mumin)\, \ZVApAV{;s}^+(g_0, a\mumin) \,,
\end{equation}
where $\zrgiVApAV{;s}^\pm(\mumin)$ is the product of the two evolution functions.

Note that $\ZtotVApAV{;s}^+ (g_0)$ depends on the renormalisation scheme $s$
only via cutoff effects. In refs.~\cite{ssf:vapav} and \cite{ssf:vapavPT}, nine
variants of Schr\"odinger functional frameworks have been explored,
each giving rise to a distinct renormalisation scheme $s$. Three of
these schemes have been deemed ``reliable'', essentially because
they display the best control of the systematic uncertainty due to 
the use of NLO perturbation theory in the calculation of $\ZtotVApAV{;s}^+$.
We refer the reader to ref.~\cite{ssf:vapav} for details; here we just
state that, following this work we label these schemes as $s = 1,3,7$. 

Given the different origin of the factors contriving to give
$\hat B_K$, the extraction of its error is somewhat
intricate. The values of $\ZtotVApAV{;s}^+ (g_0)$ for the couplings of
interest are collected in Table~\ref{tab:opZtot} of
Appendix~\ref{sec:appZ}, where a $1\%$ error is quoted. This error is
only due to the uncertainties in the determination of the lattice
renormalisation constant $\ZVApAV{;s}^+(g_0, a\mumin)$.
It is combined in quadrature with the error for the bare $B_K(g_0)$
(given in Tables~\ref{tab:respi2} and~\ref{tab:respi4})
in order to give the uncertainty for $\hat B_K$, at each $\beta$ value.
Our results for $\hat B_K$ (from $R_B^{\rm ALPHA}$), with the error
obtained as described, are gathered in Table~\ref{tab:BKcl}.
Once these results are extrapolated to the continuum
limit, a further error has to be added in quadrature, due to the
evolution function $\zrgiVApAV{;s}^\pm(\mumin)$; the reason this
is only done at the very end is that the evolution function is
a continuum quantity. As reported in ref.~\cite{ssf:vapav}, the
relative error of  $\zrgiVApAV{;s}^\pm(\mumin)$ is $2\%$.

From the results of Table~\ref{tab:BKcl} 
we see that $\hat B_K$ is essentially independent of the scheme $s$.
In the remaining analysis, we will only consider the $s=1$ results.
This arbitrary choice does not introduce any loss of generality.
\begin{table}[!t]
\begin{center}
\begin{tabular}{cccccc}
\Hline \\[-10pt]
$\beta$ & 6.0 & 6.1 & 6.2 & 6.3 & 6.45 \\ \\[-10pt]
\hline \\[-10pt]
$s = 1$ & 0.911(22) & 0.812(24) & 0.828(30) & 0.789(34) & ---  \\
$s = 3$ & 0.918(22) & 0.816(24) & 0.831(30) & 0.791(34) & ---  \\
$s = 7$ & 0.913(22) & 0.814(24) & 0.830(30) & 0.791(34) & ---  \\ \\[-10pt]
\Hline \\[-10pt]
$\beta$ & 6.0 & 6.1 & 6.2 & 6.3 & 6.45 \\ \\[-10pt]
\hline \\[-10pt]
$s = 1$ & 0.926(13) & 0.779(13) & 0.798(14) & 0.775(15) & 0.789(19)  \\
$s = 3$ & 0.933(14) & 0.783(13) & 0.800(14) & 0.777(15) & 0.789(19)  \\
$s = 7$ & 0.929(13) & 0.781(13) & 0.799(14) & 0.777(15) & 0.790(19)  \\ \\[-10pt]
\Hline
\end{tabular}
\end{center}
\caption{The value of $\hat B_K$ at fixed $\beta$-coupling, obtained from
three distinct Schr\"odinger functional renormalisation schemes $s$.
The quoted error combines in quadrature the uncertainty in the computation
of the bare $B_K$-parameter and that of the total renormalisation factor
$\ZtotVApAV{;s}^+ (g_0)$. The upper part of the table refers to $\taa= \pi/2$
results, while the lower part to $\taa= \pi/4$.}
\label{tab:BKcl}
\end{table}

Next we explore the influence of the different choices of axial
currents in the determination of $\hat B_K$. This entails comparing
results extracted from $R_B^{\rm ALPHA}$, $R_B^{\rm LANL}$ and 
$R_B^{{\rm ALPHA;~w/o}~\icA}$; see Fig.~\ref{fig:3R}.
Not surprisingly, the 
$\pi/2$ data is noisier than that of $\pi/4$. This is due to the 
fact that the latter dataset has been produced without extrapolations
in the K-meson mass. The comparison between 
$R_B^{\rm ALPHA}$ and  $R_B^{{\rm ALPHA;~w/o}~\icA}$ (which have the
same $\ZA$) shows that there are large $\icA$-related cutoff effects
at $\beta = 6.0$, which diminish as $\beta$ increases. The
$R_B^{\rm LANL}$ points fall in-between these two cases. 
Once the $\beta = 6.0$ data, afflicted by large discretisation effects,
are conservatively discarded, we are faced with three $B_K$ estimates which
display reasonable scaling behaviour.
At the largest $\beta$ values, the three $B_K$ results are
perfectly compatible.
In trying to decide which is the most reliable,
we recall that the ALPHA and LANL $\ZA$ values have been obtained for
an axial current which contains the $\icA$ counterterm. Thus,
the fact that $R_B^{{\rm ALPHA;~w/o}~\icA}$ contains the ALPHA
$\ZA$ without the $\icA$ counterterm is somewhat un-natural but 
not illicit, as it only re-introduces $\Oa$ effects in the axial current.
It could even be the source of beneficial correlators
in the $B_K$ ratio, as the corresponding $\oVApAV{O}$ counterterms are
also missing in the numerator. Nevertheless, the
$R_B^{{\rm ALPHA;~w/o}~\icA}$ results would have been on a firmer ground,
had they been obtained with an independent $\ZA$ determination 
from Ward identities, with the Clover action and without the $\icA$ term.
Also the  $R_B^{\rm LANL}$
dataset is clearly of somewhat inferior quality, as the $\ZA$
and $\icA$ are only given at a few $\beta$ values in
ref.~\cite{lat01:gupta}, while for the other $\beta$'s we had to use
simple extrapolations (cf. Table~\ref{tab:lanl}).
These general considerations place the $R_B^{\rm ALPHA}$ result on
firmer ground and justify taking the $R_B^{\rm ALPHA}$ dataset
(without the $\beta = 6.0$ point) as our best candidate for a
precision measurement of $B_K$ in the continuum limit. 

A further important element is the choice of extrapolating
function. On one hand we know that $\Oa$ effects are present and a linear
extrapolation would be theoretically justified, which however would
amplify the final error. This is especially true for our results obtained
with $\taa = \pi/2$, which have rather large errors.
On the other hand, we see that our data points show
an essentially constant behaviour in $\beta$, supporting 
a constant fit on empirical grounds, which would give smaller errors.
The procedure which is theoretically sound, while keeping 
errors under control, is a combined linear fit of the $\taa = \pi/2$ and 
$\taa = \pi/4$ results, constraining $\hat B_K$ to a unique value in
the continuum limit.
Following this procedure (see Fig.~\ref{fig:extrap_cl}) we arrive
at our final result:
\begin{equation}
\hat B_K = 0.789 \pm 0.046 \, .
\label{eq:BKbest}
\end{equation}

It is nevertheless customary to quote $B_K$ also in the $\msbar$ scheme
at a scale $\mu = 2~\GeV$. The relation between $\hat B_K$ and
$B_K^{\msbar}(\mu)$ is known at NLO (we follow the notation of~\cite{ssf:vapavPT,ssf:vapav}):
\begin{eqnarray}
\hat B_K 
&=&\Big[\dfrac{{\bar g}^2(\mu)}{4\pi}\Big]^{-\dfrac{\gamma^+_0}{2b_0}}\,\,
\exp \Big [ - \int_0^{{\bar g}(\mu)} dg \,\Big ( \dfrac{\gamma^+(g)}{\beta(g)}
- \dfrac{\gamma^+_0}{b_0 g} \Big ) \Big ] B_K^{\msbar}(\mu)
\nonumber \\
&\approx& 
\Big[\dfrac{{\bar g}^2(\mu)}{4\pi}\Big ]^{-\dfrac{\gamma^+_0}{2 b_0}}\,\,
\Big [ 1 - {\bar g}^2(\mu) 
\Big ( \dfrac{\gamma^+_1 b_0 - \gamma^+_0 b_1}{2 b_0^2} \Big ) \Big ]
B_K^{\msbar}(\mu) \,.
\label{eq:PT-RGI}
\end{eqnarray}
The gauge coupling anomalous dimension coefficients $b_0$ and $b_1$
are universal. The LO operator anomalous dimension coefficient $\gamma^+_0$
is also universal, while the NLO one $\gamma^+_1$ depends on the choice
of scheme. For $\msbar$ one also needs to specify the details
of the $\gamma_5$ matrix regularisation (i.e. DRED, NDR, HV). The values of
these coefficients are
collected in ref.~\cite{ssf:vapavPT}. Note that in the DRED scheme the
renormalised coupling we use is the one defined in the $\msbar$ scheme
(rather than the one used in the original DRED paper of
ref.~\cite{Altarelli:1980fi}). Our $\msbar$ results are
\begin{eqnarray}
B_K^{\msbar,{\rm DRED}}(2~\GeV) &=& 0.599(36) \,. \\[10pt]
B_K^{\msbar,{\rm NDR}}(2~\GeV) &=& 0.573(34) \,, \\[10pt]
B_K^{\msbar,{\rm HV}}(2~\GeV) &=& 0.633(38) \,.
\end{eqnarray}
The errors cited above arise from three uncertainties added in
quadrature: 
\begin{itemize}
\item The $\hat B_K$ error of Eq.~(\ref{eq:BKbest}).
\item The uncertainty of $\Lambda_{\msbar}$, related to the
renormalisation group running of $\bar g^2_{\msbar}$. We have used
$r_0\Lambda_{\msbar} = 0.602(48)$ from~\cite{mbar:pap1} to find $\bar g^2_{\msbar}(2\ {\rm
GeV}) = 2.54(8)$.
\item The uncertainty arising from the truncation of the perturbative series, 
when passing from the first, general expression of Eq.~(\ref{eq:PT-RGI}),
to the second, perturbative one. This is estimated by also calculating
the perturbative factor through numerical integration of the exponent
of the general expression (with $\gamma^+$ and $\beta$ truncated to the
highest available order in PT). 
We take the spread of these results as our estimate of the $O(g^4)$
uncertainties.
\end{itemize}
 
\section{Conclusions}
\label{sec:concl}

In the present work we have presented a $B_K$ value obtained
from first principles, without any
uncontrolled approximations, except for the quenching of the sea quarks
and the use of degenerate strange and down valence quarks.
The only input from experiment has been the setting of the scales from
the physical K-meson mass and the $r_0$ parameter. 

Both matrix element renormalisation and RG-running are non-perturbative.
NLO perturbation theory has only been used at very high scales, 
i.e.~deep in the perturbative regime.
The bare $B_K$
has been computed in the lattice regularisation with Wilson fermions,
using two variants of tmQCD, at several couplings. Thus, a well
controlled continuum limit extrapolation was possible, 
in which we have conservatively
dropped the result of the coarsest lattice. This was imposed by the
large cutoff effects which, at small $\beta$ values, arise from different
choices of the $\Oa$ improvement term proportional to $c_{\rm A}$ 
of the axial current.
This source of systematic error, as well as $\Oa$ cutoff effects in
general, may be removed in future simulations performed
along the lines proposed in ref.~\cite{FrezzoRoss2}. In any case, our
result has been obtained with a good control of all sources
of uncertainty, except for the quenching of the fermion determinant.
Most recent quenched results, derived with different lattice
regularisations, usually with a poorer control of systematic errors,
are in agreement with our value~\cite{Dawson:lat05}.
It is interesting that this is not the case for
the most recent measurement of $B_K$ with Wilson fermions~\cite{SPQR:bk}. 
This is discussed in detail in Appendix~\ref{sec:appcomp}.
 
\section*{Acknowledgments}
We are grateful to I.~Wetzorke and the ZeRo collaboration for
providing us with the $\kappa_{\rm crit}(\beta = 6.3)$ estimate.
We also wish to thank M.~Golterman, M.~Guagnelli, V.~Lubicz,
M.~Papinutto and R.~Sommer for discussions.
F.P. acknowledges financial support from the
Alexander-von-Humboldt Stiftung.
Finally, we thank CERN, DESY, DESY-Zeuthen, INFN-Rome2 and
the Univ. Aut\'onoma de Madrid for providing hospitality to several
members of our collaboration at various stages of this work,
as well as the DESY-Zeuthen computing centre for its support.

\begin{appendix}
\section{Renormalisation constants and mixing coefficients}
\label{sec:appZ}

In this appendix we collect all the results we have used from
other sources, concerning bare parameters, renormalisation constants,
$\Oa$ coefficients, matching factors etc.

Our simulations were produced with the Sheikholeslami-Wohlert action, using the
non-perturbative value of the Clover term given in ref.~\cite{impr:pap3}:
\begin{equation}
\icsw = \dfrac{1 - 0.656 g_0^2 - 0.512 g_0^4 - 0.054 g_0^6}
{1 - 0.922 g_0^2} \,\,, \qquad 0 \le g_0^2 \le 1 \,\,.
\end{equation}

For the tuning of the quark masses, as described in subsect.~\ref{sec:masstune},
we need the values of the critical hopping parameters. These may be found in various
sources, as shown in
Table~\ref{tab:kcrit}.
\vskip 0.2cm
\begin{table}[!ht]
\begin{center}
\begin{tabular}{ccccccc}
\Hline \\[-10pt] 
$\beta$ & 6.0 & 6.1 & 6.2 & 6.3 & 6.4 & 6.45 \\ \\[-10pt]
\hline \\[-10pt]
$\kappa_{\rm crit}$ & 0.135196 & 0.135496 & 0.135795 & 0.135823 &  0.135720 &
0.135701 \\ \\[-10pt]
\hline \\[-10pt]
ref. & \cite{impr:pap3} & \cite{mbar:charm1} & \cite{impr:pap3} & ZeRo
Coll. &  
\cite{impr:pap3} & \cite{mbar:charm1} \\ \\[-10pt]
\Hline
\end{tabular}
\end{center}
\caption{The critical hopping parameter $\kappa_{\rm crit}$ at several $\beta$-values.}
\label{tab:kcrit}
\end{table}
We also need the non-perturbative expressions for the renormalisation constant
$Z\equiv Z_{\rm P}/(Z_{\rm S} Z_{\rm A})$ of ref.~\cite{impr:babp},
\begin{equation}
Z= (1 + 0.090514 g_0^2)\,\dfrac{(1-0.9678 g_0^2 + 0.04284 g_0^4 - 0.04373 g_0^6)}
{(1-0.9678 g_0^2)}\biggr|_{0 \le g_0^2 \le 1} \,\,,
\end{equation}
as well as the axial current normalisation $Z_{\rm A}$, which will be discussed below. At present,
it suffices to say that we use the ALPHA collaboration estimate of $Z_{\rm A}$ for the quark
mass tuning. Finally, for the mass $\Oa$ counterterms $b_{\rm m}$ and $b_\mu$ we use the
non-perturbative and
1-loop perturbative results of refs.~\cite{impr:babp} and \cite{tmqcd:pap2} respectively:
\begin{eqnarray}
b_{\rm m} &=& (-0.5 - 0.09623 g_0^2)\dfrac{(1-0.6905 g_0^2 + 0.0584 g_0^4)}
{(1-0.6905 g_0^2)}\biggr|_{0 \le g_0^2 \le 1} \,\,,
\\[8pt]
b_\mu &=& - 0.103 \,\, C_F \,\, g_0^2 \,\,,
\end{eqnarray} 
with $C_F = (N_c^2 -1)/(2N_c)$ and $N_c = 3$ the number of colours. 
The last expression is obtained once the arbitrary value of $\tilde b_{\rm m}$
has been fixed to $\tilde b_{\rm m} = -1/2$ (see ref.~\cite{tmqcd:pap2} for a detailed
explanation). 

The non-perturbative expressions for the current normalisation constants
$Z_{\rm A}$ and $Z_{\rm V}$ have been computed by the ALPHA
collaboration in ref.~\cite{impr:pap4}:
\begin{eqnarray}
Z_{\rm A} &=& \dfrac{1 - 0.8496 g_0^2 + 0.0610 g_0^4}{1 - 0.7332 g_0^2} \,\,,
\qquad 0 \le g_0^2 \le 1 \,\,,
\label{eq:ZA}
\\[8pt]
Z_{\rm V} &=& \dfrac{1 - 0.7663 g_0^2 + 0.0488 g_0^4}{1 - 0.6369 g_0^2} \,\,,
\qquad 0 \le g_0^2 \le 1 \,\,.
\label{eq:ZV}
\end{eqnarray} 
with a precision which is better than $0.6\%$ and  $0.4\%$ respectively.
The same collaboration finds, for the improvement coefficient of the
axial current~\cite{impr:pap3}:
\begin{equation}
\icA = - 0.00756 g_0^2 \,\,\, \dfrac{1 - 0.748 g_0^2} {1 - 0.977 g_0^2} \,\,,
\qquad 0 \le g_0^2 \le 1 \,\,.
\end{equation}
For the same quantities, the LANL collaboration quote in ref.~\cite{lat01:gupta} 
as their best results those reproduced here in Table~\ref{tab:lanl}.
In that work, only results at three $\beta$-values have been computed.
We have interpolated/extrapolated linearly these results (with the two 
independent errors of $Z_{\rm A}$ added in quadrature) in order to obtain
estimates for all $\beta$-values used in this work.
\vskip 0.2cm
\begin{table}[!ht]
\begin{center}\scriptsize
\begin{tabular}{ccccccc}
\Hline \\[-6pt] 
$\beta$ & $6.0$ & $6.1^\ast$ & $6.2$ & $6.3^\ast$ & $6.4$ &
$6.45^\ast$ \\ \\[-6pt]
\hline \\[-6pt]
$Z_{\rm V}$ & 0.770(1) & 0.7791(5) & 0.7874(4) &  0.7951(5) &  0.802(1) &
0.8070(9) \\ \\[-6pt]
\hline \\[-6pt]
$Z_{\rm A}$ & 0.802(2)(8) & 0.809(1)(5) & 0.815(2)(5) &  0.818(1)(3) &  0.822(1)(4) & 
0.825(1)(5) \\ \\[-6pt]
\hline \\[-6pt]
$\icA$ & $-0.038(4)$ & $-0.036(2)$ & $-0.033(3)$ &  $-0.033(2)$ &  $-0.032(3)$ & 
$-0.031(3)$ \\ \\[-6pt]
\Hline
\end{tabular}
\end{center}
\caption{Current normalisation factors $Z_{\rm V}$, $Z_{\rm A}$ and axial current improvement
coefficient $\icA$, as reported in ref.~\cite{lat01:gupta} at $\beta = 6.0,6.2,6.4$.
For the other $\beta$-values, indicated by an asterisk, we have obtained results
by linear interpolation/extrapolation.}
\label{tab:lanl}
\end{table}
Upon using $Z_{\rm A}$ and $Z_{\rm V}$ together for the composition of the tmQCD vector
and axial currents, we account for their correlations by conservatively increasing
their error to a $2\%$ and $0.5\%$ uncertainty respectively.

As explained in sect.~\ref{sec:SFtmQCD}, in the computation of $B_K$ the currents we use
are normalised with $Z_{\rm A}$ and $Z_{\rm V}$ and only partially improved by
the counterterm $\icA$.
The computation of the effective masses $a M_{\rm PS}^{\rm eff}$ (and eventually the 
improved decay constant $F_K$) is done using fully improved currents, including the
counterterms proportional to the quark mass. For the vector current, the counterterm 
$\ibV$ has been computed non-perturbatively in~refs.\cite{impr:pap4,impr:pap5}:
\begin{equation}
\ibV = \dfrac{1 - 0.7613 g_0^2 + 0.0012 g_0^4 - 0.1136 g_0^6}
{1 - 0.9145 g_0^2} \,\,, \qquad 0 \le g_0^2 \le 1 \,\,,
\end{equation}
while the corresponding coefficient $\ibA$ of the axial current
is given in 1-loop perturbation theory in~ref.\cite{impr:pap5}:
\begin{equation}
\ibA = 1 + 0.11414 \,\, C_F \,\, g_0^2 \,\,.
\end{equation}
The twisted mass counterterms for the currents also depend on the
arbitrariness of $\tilde b_{\rm m}$; setting again  $\tilde b_{\rm m} = -1/2$,
ref.~\cite{tmqcd:pap2} gives at 1-loop in the perturbative expansion:
\begin{eqnarray}
\ibAtil = 1 + 0.086 \,\, C_F \,\, g_0^2 \,\,,
\\[10pt] 
\ibVtil = 1 + 0.074 \,\, C_F \,\, g_0^2 \,\,.
\end{eqnarray}
The relation between the lattice spacing and the scale $r_0$ is given at
ref.~\cite{pot:intermed}:
\begin{equation}
\ln(a/r_0) = - 1.6804 - 1.7331 (\beta - 6) + 0.7849 (\beta - 6)^2
- 0.4428 (\beta - 6)^3 \,\,.
\end{equation}
The total renormalisation factor $\ZtotVApAV{;s}^+ (g_0)$, which connects the bare 
$B_K$ to its RGI value $\hat B_K$, has been computed non-perturbatively in ref.~\cite{ssf:vapav}.
In table~\ref{tab:opZtot} we collect these results for the couplings of interest.
\begin{table}[!ht]
\begin{center}
\begin{tabular}{ccccccc}
\Hline \\[-10pt] 
$\beta$ & $6.0$ & $6.1$ & $6.2$ & $6.3$ & $6.4$ & $6.45$ \\ \\[-10pt]
\hline \\[-10pt]
$\ZtotVApAV{;1}^+$ & 0.884 & 0.901 & 0.918 &  0.935 &  0.952 & 0.960
\\ \\[-10pt]
\hline \\[-10pt]
$\ZtotVApAV{;3}^+$ & 0.890 & 0.905 & 0.921 &  0.937 &  0.953 & 0.962 \\ \\[-10pt]
\hline \\[-10pt]
$\ZtotVApAV{;7}^+$ & 0.886 & 0.903 & 0.920 &  0.937 &  0.954 & 0.962 \\ \\[-10pt]
\Hline
\end{tabular}
\end{center}
\caption{Operator total renormalisation factors $\ZtotVApAV{;s}^+ (g_0)$ for three
Schr\"odinger functional schemes ($s = 1,3,7$). The error of these estimates is at most $1\%$.}
\label{tab:opZtot}
\end{table}

\section{Improvement of quenched quark bilinear operators}
\label{sec:appimp}

In the quenched approximation, the improvement of two-fermion
composite operators
has been carried out in \cite{tmqcd:pap2} for tmQCD
with a degenerate isospin doublet of twisted flavours.
Here we adapt these results for the cases of interest.
The most efficient way to do this is by first considering a
generalisation of tmQCD, given by the action
\begin{equation}
 \label{tmQCD_actiongen}
 S_F = a^4 \sum_x \,\, \bar{\psi}(x) [\Dwd 
+ {\bf m} + i \ci{\mu} \gamma_5 ]\psi(x) \,,
\end{equation}
where $\Dwd$ is the Wilson-Dirac operator,
the quark spinor $\psi$ contains in general $N_f$ flavours,
${\bf m}$ and $\ci{\mu}$ are diagonal mass matrices in flavour
space and ${\bf m_q}$ the subtracted mass matrix. The two actions
considered in our work are obtained for $N_f=3$ with
${\bf m_q} = \diag(0, 0, m_{q,s}); \ci{\mu} = \diag(\mu_l, -\mu_l, 0)$
(for twist angle $\pi/2$) and $N_f = 2$ with
${\bf m_q} = \diag(m_q, m_q); \ci{\mu} = \diag(\mu_l, -\mu_l)$
(for twist angle $\pi/4$).

The improvement pattern of a quark bilinear operator
$O_{ij} = \bar \psi_i \Gamma \psi_j$ (with $i,j$ denoting distinct 
flavours and $\Gamma$ any Dirac matrix), is established with
the aid of discrete symmetries of the theory, namely
charge conjugation combined with flavour exchange
$m_i \leftrightarrow m_j$, $\mu_i \leftrightarrow \mu_j$,
parity combined with sign flips of the twisted masses
$\mu_i \rightarrow -\mu_i$ and time reversal
combined with sign flips of the twisted masses.
Using these symmetries we obtain for on-shell improvement
\begin{align}
  (S_{\rm R})_{ij} &= Z_{\rm S} [1 + \ibS a \dfrac{m_{q,i}+m_{q,j}}{2}]
    [S_{ij} - i \ibStil a \dfrac{\mu_i+\mu_j}{2} P_{ij}] \,,
\label{eq:impSij} \\[8pt]
  (P_{\rm R})_{ij} &= Z_{\rm P} [1 + \ibP a \dfrac{m_{q,i}+m_{q,j}}{2}]
    [P_{ij} - i \ibPtil a \dfrac{\mu_i+\mu_j}{2} S_{ij}] \,,
\label{eq:impPij} \\[8pt]
  (V_{\rm R})_{\mu,ij} &= Z_{\rm V} [1 + \ibV a \dfrac{m_{q,i}+m_{q,j}}{2}]
    [V_{\mu,ij} - i \ibVtil a \dfrac{\mu_i-\mu_j}{2} A_{\mu,ij}
+ a \icV \tilde \partial_\nu T_{\mu\nu}] \,,
\label{eq:impVji} \\[8pt]
  (A_{\rm R})_{\mu,ij} &= Z_{\rm A} [1 + \ibA a \dfrac{m_{q,i}+m_{q,j}}{2}]
    [A_{\mu,ij} - i \ibAtil a \dfrac{\mu_i-\mu_j}{2} V_{\mu,ij}
+ a \icA \tilde \partial_\mu P] \,,
\label{eq:impAij} \\[8pt]
  (T_{\rm R})_{\mu\nu,ij} &= Z_{\rm T} [1 + \ibT a \dfrac{m_{q,i}+m_{q,j}}{2}]
    [T_{\mu\nu,ij} - i \ibTtil a \dfrac{\mu_i+\mu_j}{2} 
\tilde T_{\mu\nu,ij}
+ a \icT \tilde \partial_\mu V_{\nu,ij} ] \,,
\label{eq:impTij}
\end{align}
where $\tilde \partial$ is a lattice symmetrised derivative, and
$\tilde T_{\mu\nu} = \half\epsilon_{\mu\nu\rho\sigma}T_{\rho\sigma}$.
The coefficient signs are chosen so as to agree with the results
of ref.~\cite{tmqcd:pap2}

\section{Linear and chiral log extrapolations}
\label{sec:applinfit}

\begin{figure}
\begin{center}
\epsfig{figure=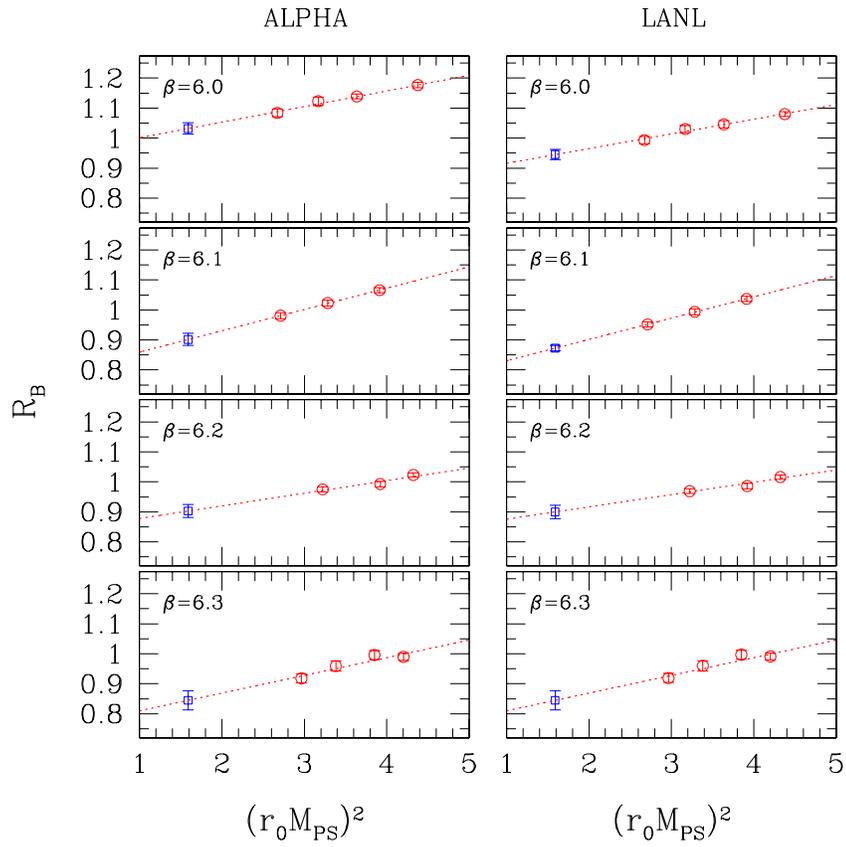, width=12.5 true cm}
\vskip -0.5cm
\caption{Linear extrapolation of $R_{\rm B}$ to the physical kaon mass
at $\taa = \pi/2$.}
\label{fig:extrap_pi2}
\end{center}
\end{figure}

Here we address the problem of whether it is justified,
in the $\pi/2$ tmQCD formalism, to compute $B_K$ by
performing the linear extrapolations of~\refig{fig:extrap_pi2}
from high $(r_0 M_{\rm PS})^2$ values to the physical point.
An alternative functional behaviour could be that of
eq.~(\ref{eq:BKchPT}), which in the degenerate flavour case becomes
\begin{equation}
B_K = B \Big [ 1 - 3y \ln y + b y \Big ] \, .
\label{eq:BKchPT-den}
\end{equation}
This is a continuum limit expression. As it is valid
close to the chiral limit, there is no {\it a priori} reason
for it to be used in the mass range we are simulating.
However, its validity can be easily tested on our data.

We have set three reference values of $y$, spanning
the data points of our $\pi/2$ simulations. At fixed $\beta$, we have
computed $B_K(\beta,y)$ for these three reference $y$-values,
by linear interpolation (or short extrapolation). We then compute
the RGI quantity $\hat B_K(\beta,y)$ (as explained in sect.~\ref{sec:bkcon})
at $\beta = 6.1, 6.2, 6.3$, for which we are confident that scaling
has set in. Finally, we extrapolate $\hat B_K(\beta,y)$ linearly to the
continuum limit, thus obtaining three continuum $\hat B_K(y)$
estimates, where \req{eq:BKchPT-den} can be applied.

\begin{figure}
\begin{center}
\epsfig{figure=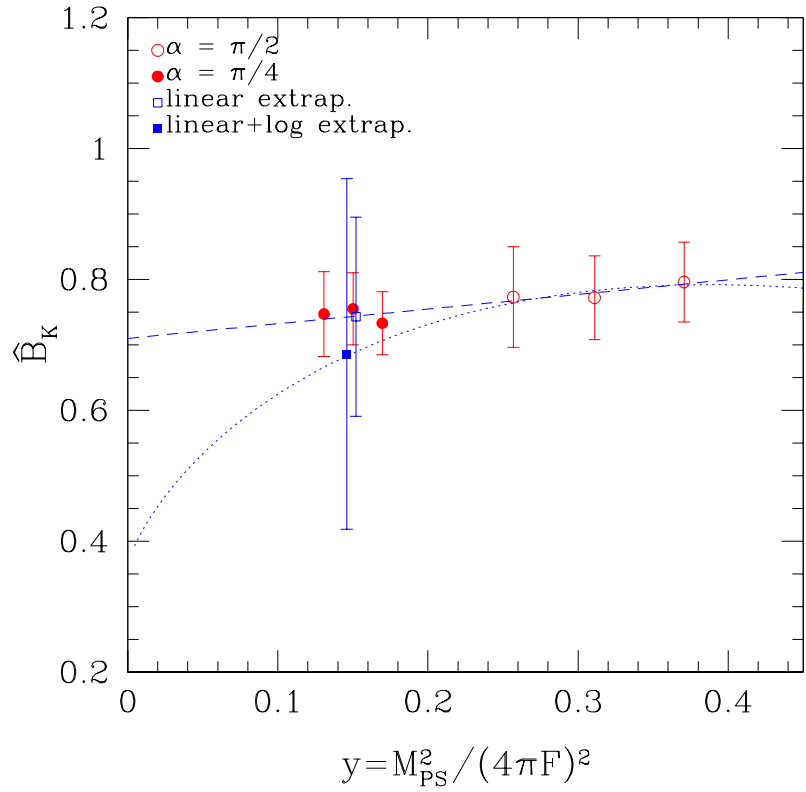, width=11.2 true cm}
\caption{Linear vs. chiral log extrapolation of $\hat B_K$ to its
physical mass value.}
\label{fig:chPT}
\end{center}
\end{figure}

Next we perform both a linear extrapolation and one based on
\req{eq:BKchPT-den}. The results of these two extrapolated $\hat B_K$
values are shown in Fig.~\ref{fig:chPT}, where they are compared to the
continuum $\hat B_K$ estimates, obtained with the same procedure from the
$\pi/4$ data, practically computed at the physical mass point.
We see that, within large errors, the chiral logs are not
resolved by our data. Nevertheless, it appears that 
the linear extrapolation performs somewhat better than the chiral log one.

A last interesting point concerns the comparison of our best
$\hat B_K$ estimate (see \req{eq:BKbest}) to the one obtained
in this Appendix, from the linear extrapolation:
\begin{equation}
\hat B_K = 0.74 \pm 0.15 \,.
\label{eq:BKapp}
\end{equation}
The difference in the extraction of the two results is essentially 
that the extrapolations in the mass and the continuum limit
have been swapped~\footnote{Another important difference is that the
value of \req{eq:BKbest} is the result of a combined fit of the
$\pi/2$ and $\pi/4$ data, while that of \req{eq:BKapp} involves
only the long extrapolation of the $\pi/2$ data. Thus, the errors
of the latter estimate are larger.}. The fact that the two results
are compatible means that, for the mass ranges under consideration, 
the order in which these limits are taken is irrelevant. This is
clearly not true close to the chiral limit.

\section{Exceptional configurations}
\label{sec:excconf}

The initial motivation behind the introduction of tmQCD was the problem of exceptional 
configurations~\cite{wupp:latt86}, which is solved by cutting-off in the IR limit the low-lying 
eigenvalues of the lattice Dirac operator, with the twisted mass $\mu_l$ \cite{tmqcd:pap1}.
The tmQCD action we have adopted so far refers to the light flavours,
whereas the strange quark propagator, 
being regularised by the standard Wilson fermion action, is not immune from exceptional
configurations. The common lore~\cite{mbar:pap3} is that for the values of $\kappa_s$
of Table~\ref{tab:runspi2} (corresponding to a $K$-meson of more than $600~\MeV$), no exceptional
configurations ought to occur. In fact, none of the runs reported above showed any signs of
such a problem.
\begin{figure}
\begin{center}
\epsfig{figure=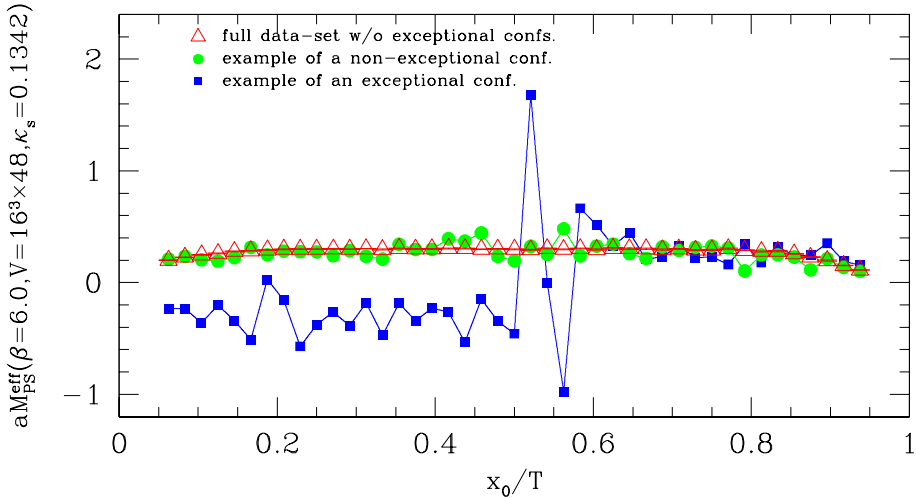, width=13 true cm}
\vspace{-60 truemm}
\caption{The value of $a M_{\rm PS}^{\rm eff}$ for: (i) an exceptional configuration; (ii) a 
typical configuration of our ensemble; (iii) the mean value,
averaged on the configuration ensemble, without the exceptional configuration.}
\label{fig:except}
\end{center}
\end{figure}
Nevertheless, evidence of an exceptional configuration showed up in an aside simulation, 
after 363 measurements at $\beta = 6.0$, a $16^3 \times 48$ lattice volume and 
$\kappa_s = 0.1342$ (for the record we mention that in the run, $\kappa_d = 0.135194$ and
$a\mu_d = 0.01051$). Thus the problem arose, somewhat surprisingly,
at a $\kappa_s$ believed to be safe, on empirical grounds.

In this respect we have considered the effective mass
$a M_{\rm PS}^{\rm eff}$, as obtained in standard (non-twisted)
lattice QCD. This is like the quantity computed in \req{eq:effmass}, 
without the $f_{\rm V}$ correlation functions and with $f_{\rm A}$ made up of only
standard (non-twisted) Wilson quarks (i.e. the strange
flavours of our formulation). In Fig.~\ref{fig:except} 
we plot $a M_{\rm PS}^{\rm eff}$, calculated at the exceptional configuration as well as
at a typical configuration of our ensemble. These results are compared to the average 
$a M_{\rm PS}^{\rm eff}$, computed on the whole ensemble, save for the exceptional configuration.
This plot is a clear warning against simulating even at reputedly safe values
of the Wilson hopping parameter. We point out that several lattice measurements, besides our own,
have been performed by other collaborations at this hopping parameter (e.g. see ref.~\cite{SPQR:bk}).

\section{Comparison with the $B_K$ computation of ref.~\cite{SPQR:bk}}
\label{sec:appcomp}

In this Appendix we compare our results to the most
recent ones obtained with Wilson fermions~\cite{SPQR:bk},
where $\hat B_K = 0.96 \pm 0.10$ is quoted. There are a few
similarities and several differences between the two approaches:
\begin{itemize}
\item In both approaches Wilson fermions are used, but in the
present work we have a twisted term in the lattice action.
In ref.~\cite{SPQR:bk} the $\Delta S = 2$ bare matrix element
is computed on periodic lattices, from a four-point correlation
function of the operator $O_{\rm VA+AV}$ (which is related to the
three-point correlation function of $O_{\rm VV+AA}$ through Ward
identities). In the present work it is computed from a three-point
correlation function of the operator $O_{\rm VA+AV}$ (which is related
to the three-point correlation function of $O_{\rm VV+AA}$ through the tmQCD
formalism). Schr\"odinger functional boundary conditions are used,
which amounts to smeared pseudoscalar sources at the two time boundaries.
\item In both approaches matrix elements are renormalised non-perturbatively.
In ref.~\cite{SPQR:bk} the renormalisation constants are computed in the 
RI/MOM scheme, at scales of a few GeV. The RGI $\hat B_K$ requires
also the operator RG-running; this is performed in NLO perturbation
theory. In the present work, matrix elements are renormalised at
hadronic scales in a SF scheme. The RG-running is performed
non-perturbatively, using finite volume techniques, up to a few
tens of GeV and subsequently with NLO perturbation theory.
\item In the present work, $B_K$ is extracted directly from
suitable ratios of correlation functions. In ref.~\cite{SPQR:bk},
$B_K$ is obtained from the slope of the four-fermion operator matrix
element, as a function of its vacuum saturation estimate $(8/3)F_K^2 m_K^2$.
This linear behaviour is predicted by lowest order chiral perturbation
theory and is also seen in the data produced at pseudoscalar masses above
the value of the physical $K$-meson mass.
\end{itemize}

\begin{figure}[!h]
\begin{center}
\epsfig{figure=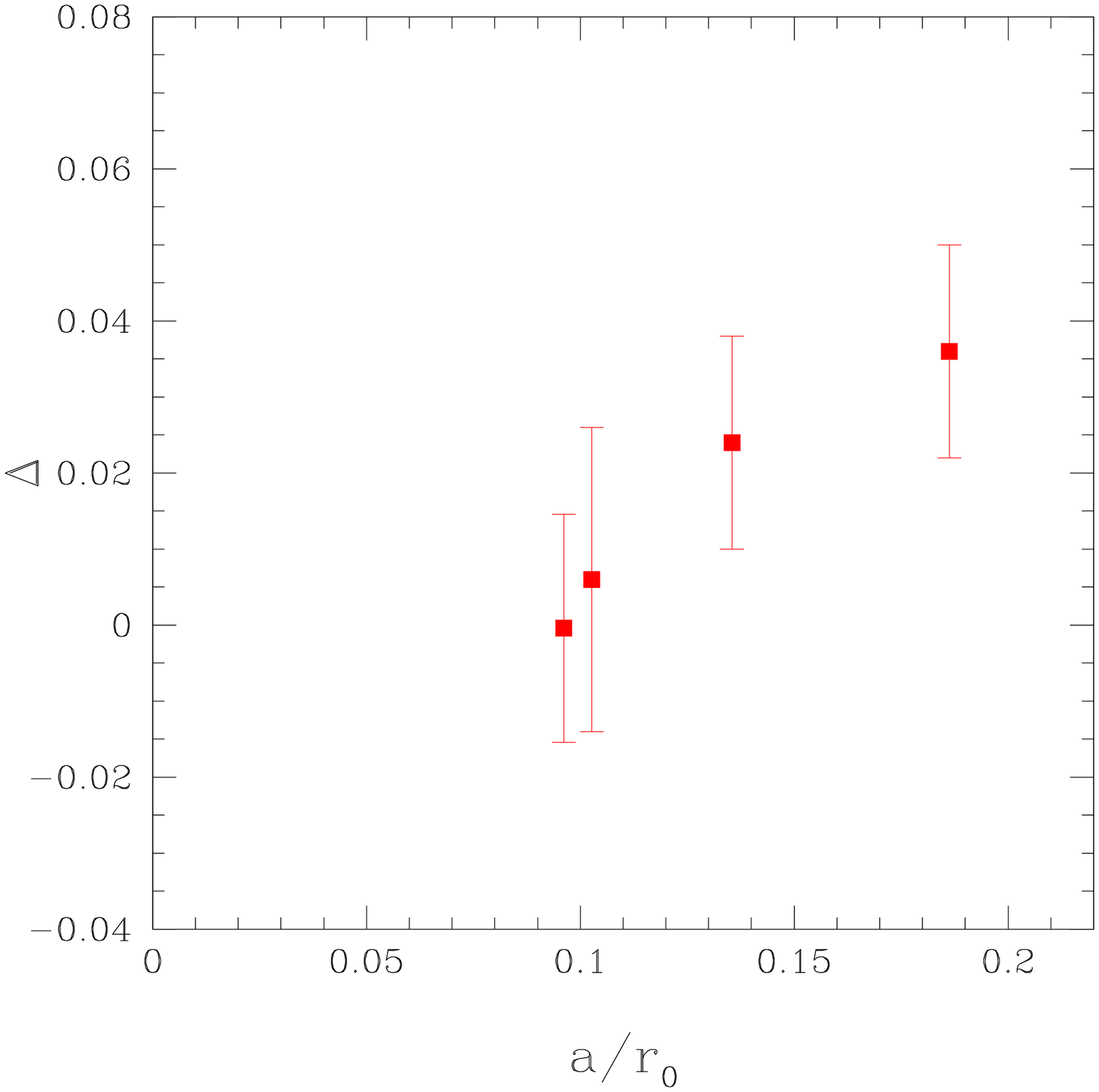, width=8.0 truecm}
\epsfig{figure=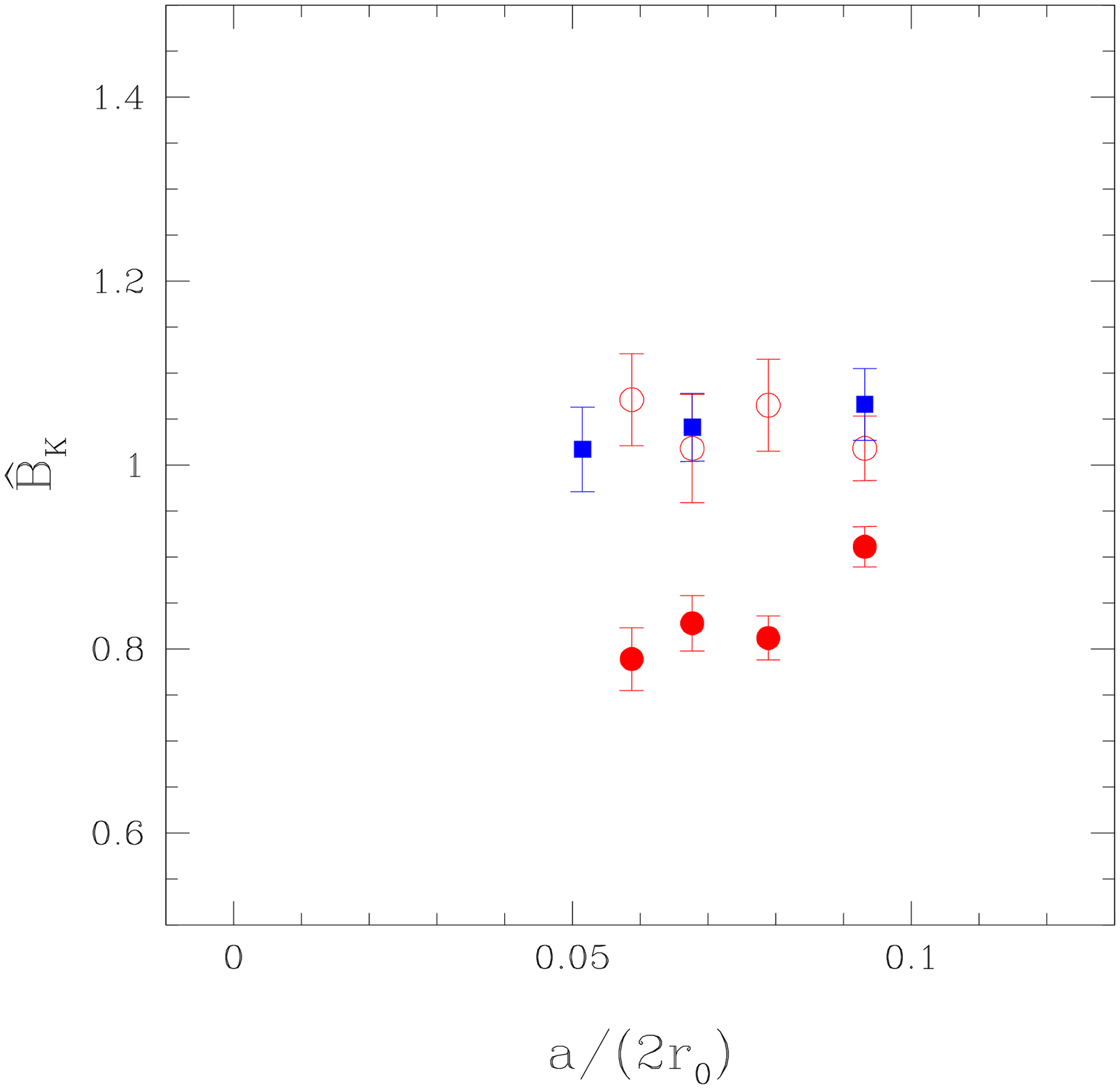, width=8.0 truecm}
\caption{Top: $\Delta$ is the difference of $\ZtotVApAV{;1}^+$, computed in
the SF scheme~\cite{ssf:vapav} and $\ZtotVApAV{;{\rm RI/MOM}}^+$,
computed in the RI/MOM scheme~\cite{Zrimom:clov}, normalised
by $\ZtotVApAV{;1}^+$. It is a discretisation effect which vanishes
close to the continuum limit. Bottom: Comparison of $\hat B_K$
obtained from three different methods: (i) results of
ref.~\cite{SPQR:bk}  (filled squares); (ii) results obtained 
from our data (tmQCD and SF-renormalisation), 
using the method of ref.~\cite{SPQR:bk}
(open circles); (iii) the main results of the present work
(filled circles).}
\label{fig:Alpha-SPQR}
\end{center}
\end{figure}

It is interesting to try and understand the source of the difference
in the $\hat B_K$ value quoted by us and by the authors of ref.~\cite{SPQR:bk}.
First we compare the quantity  $\ZtotVApAV{;1}^+ (g_0)$ of
ref.~\cite{ssf:vapav} to the same quantity, calculated in the
the RI/MOM scheme; this is done by combining the $Z_{\rm VA+AV}$ values
of ref.~\cite{Zrimom:clov}, used in the $B_K$ computation of
ref.~\cite{SPQR:bk}, with the NLO running quoted in
refs.~\cite{rm1:4ferm-nlo,mu:4ferm-nlo}. The two estimates do not
depend on the renormalisation scheme in which they have been computed.
They only differ by (i) discretisation effects; (ii) the reliability
(convergence) of the NLO perturbative running, calculated in two
different schemes and applied at two different scales.
In Fig.~\ref{fig:Alpha-SPQR} we show the quantity
\begin{equation}
\Delta = \dfrac{\ZtotVApAV{;1}^+ - \ZtotVApAV{;{\rm RI/MOM}}^+}{\ZtotVApAV{;1}^+}
\end{equation}
as a function of the lattice spacing.
It is clear that a barely significant difference at the largest
lattice spacing, quickly disappears as the continuum limit is
approached. Thus it seems that renormalisation is not responsible
for the different $\hat B_K$ results quoted by the two groups.

In order to have as direct a comparison as possible, we compute
in the $\taa = \pi/2$ case the quantity
\begin{equation}
\hat R^\prime = \dfrac{ i Z_{\rm VA+AV} F_{\rm VA+AV} }
{2 [ \fP(x_0) - i \dfrac{Z_{\rm S}}{Z_{\rm P}} \fS(x_0) ]
 [ \fP^\prime(T-x_0) - i \dfrac{Z_{\rm S}}{Z_{\rm P}} \fS^\prime(T-x_0) ] } \,.
\label{eq:Rprime}
\end{equation}
The computation is performed at each value of the inverse coupling
$\beta$ and mass parameters $\kappa_s$ and ($\kappa_d, a\mu$) of
Table~\ref{tab:runspi2}. The normalisation factor
$Z_{\rm S}/Z_{\rm P} = (Z Z_{\rm A})^{-1}$ is calculated from the 
$Z$ and $Z_{\rm A}$ ALPHA collaboration estimates quoted
in Appendix~\ref{sec:appZ}.

A glance at Eqs.~(\ref{eq:Sus_transf},\ref{eq:Pus_transf},
\ref{eq:impSij},\ref{eq:impPij}) shows that the above ratio
tends, at large time separations, to the quantity
\begin{equation}
\hat R^\prime \rightarrow 
\dfrac{ Z_{\rm VA+AV} \langle \bar K^0 \vert O_{\rm VA+AV} \vert K^0 \rangle}
{\vert \langle 0 \vert P_{sd} \vert K^0 \rangle \vert^2} \,.
\end{equation}
Finally, for each set of coupling and mass parameter values,
we construct the quantity 
\begin{equation}
\hat R^\prime_{\rm RGI} = \ZtotVApAV{;1}^+ (g_0) \,\, R^\prime
\label{eq:Rrgi}
\end{equation}
in exactly the same way we have computed $\hat B_K$ of
eq.~(\ref{eq:BKcont}). This quantity may be directly compared to the
ratio $\hat R^{\rm w/o \,\,\, subtr.}$ of ref.~\cite{SPQR:bk} (see
Table 1 of that work; the same quantity is sometimes referred to as 
$R^{rgi}$ by the authors).\footnote{Note that: (i) the suffix RGI is somewhat
misleading, as it only refers to the numerator of $R^\prime_{\rm RGI}$,
while the denominator is a bare quantity; (ii) following ref.~\cite{SPQR:bk},
the matrix elements of the denominator of $\hat R^\prime$ are
not improved.}

We also compute the quantity
\begin{equation}
X^\prime = 
\dfrac{8}{3}\dfrac{[ Z_{\rm A} \fA(x_0) - i Z_{\rm V} \fV(x_0) ]
[ Z_{\rm A} \fP^\prime(T-x_0) - i Z_{\rm V} \fS^\prime(T-x_0) ] }
{[ \fP(x_0) - i \dfrac{Z_{\rm S}}{Z_{\rm P}} \fS(x_0) ]
 [ \fP^\prime(T-x_0) - i \dfrac{Z_{\rm S}}{Z_{\rm P}} \fS^\prime(T-x_0) ] }\,.
\label{eq:Xprime}
\end{equation}
This ratio tends, at large time separations, to the quantity
\begin{equation}
X^\prime \rightarrow 
\dfrac{8}{3} \dfrac{F_K^2 m_K^2}
{\vert \langle 0 \vert P_{sd} \vert K^0 \rangle \vert^2}\,,
\end{equation}
which is the same quantity as $X$ of ref.~\cite{SPQR:bk}.
For both $\hat R^\prime_{\rm RGI}$ and $X^\prime$ we have used
ALPHA collaboration renormalisation constants.

At fixed gauge coupling $\beta$, the quantity
$\hat R^\prime_{\rm RGI}$ is fit linearly as a function of
$X^\prime$; the slope is the $\hat B_K$ estimate
according to lowest order chiral perturbation theory.
In Fig.~\ref{fig:Alpha-SPQR} we compare the results
of ref.~\cite{SPQR:bk} to our results (obtained
as described here from the ratio $\hat R^\prime_{\rm RGI}$) and find
them compatible. We also see that our preferred $B_K$ estimates
(obtained on the same
gauge configurations, but extracted from the ratio $R_B$) give
a consistently lower $B_K$ estimate. This shows that the discrepancy
of the two results is not due to different regularisation and
renormalisation systematics, but due to the different method of
extraction of $B_K$.

The errors we quote for the results of the present method are larger 
than those of ref.~\cite{SPQR:bk}. This is due to the
fact that the latter results have been obtained, at fixed $\beta$ and
for several values of the hopping parameters, on the same configuration
ensemble, while we have generated a separate ensemble at each $\kappa$.
Thus, the extrapolations in the pseudoscalar mass in ref.~\cite{SPQR:bk}
benefit from the correlations between measurements. On the other
hand, for both datasets, $\hat B_K$ has been computed from the slope of
$\hat R^\prime_{\rm RGI}$ as a function of $X^\prime$. This fitting
procedure enlarges the errors, as can be seen by comparing them to
those of our main method (based upon computing $R_B$ at several
$\kappa$-values and extrapolating it to the physical Kaon mass).

\end{appendix}
\bibliography{lattice} 
\begin{figure}
\begin{center}
\epsfig{figure=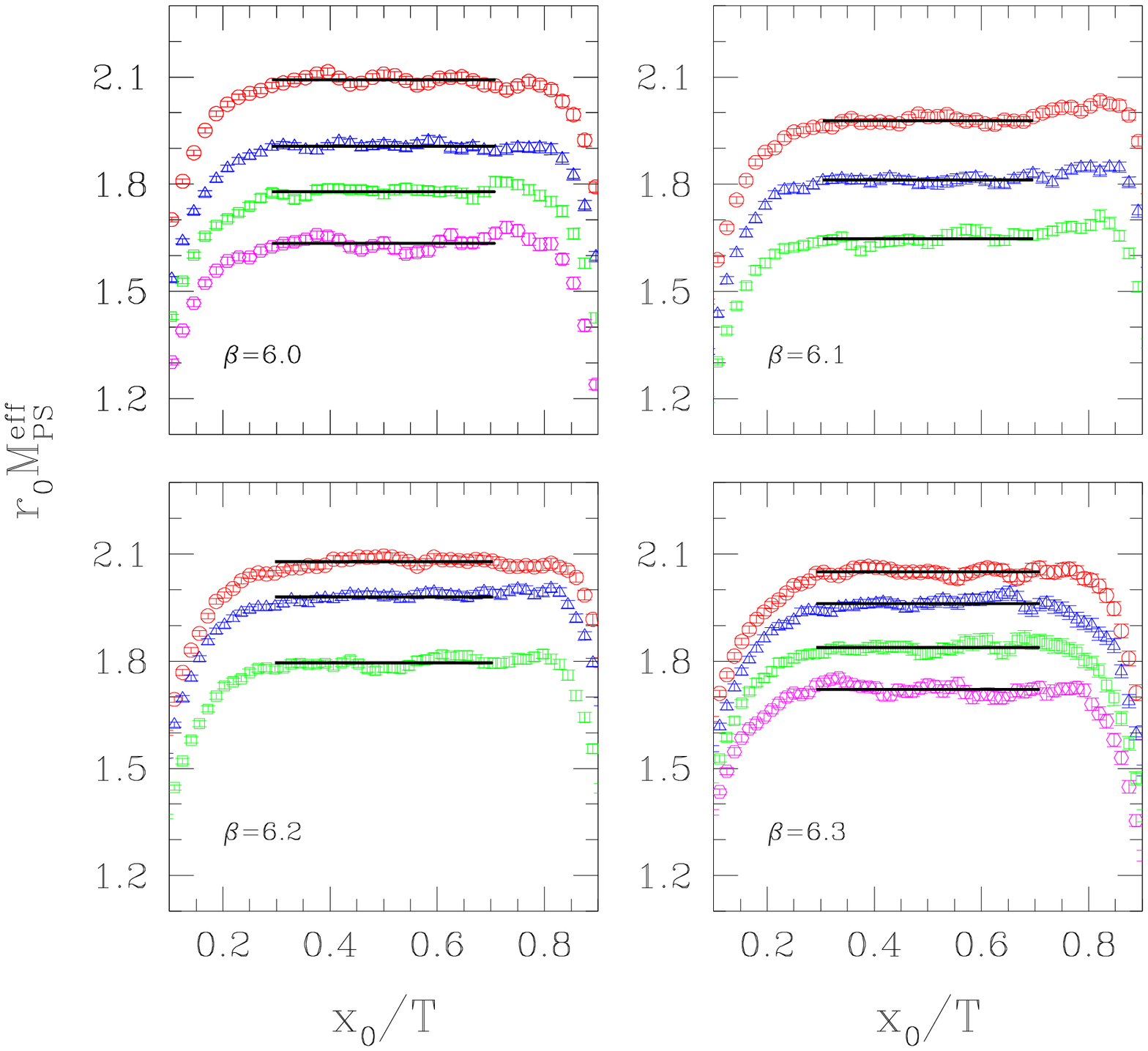, width=12.5 true cm}
\vskip -0.0cm
\caption{Plateaux for the extraction of $M_{\rm PS}^{\rm eff}$ at
$\alpha=\pi/2$. The time-range and value of each plateau is indicated
by a straight line segment.}
\label{fig:plateaux_m_pi2_1}
\end{center}
\end{figure}

\begin{figure}
\begin{center}
\epsfig{figure=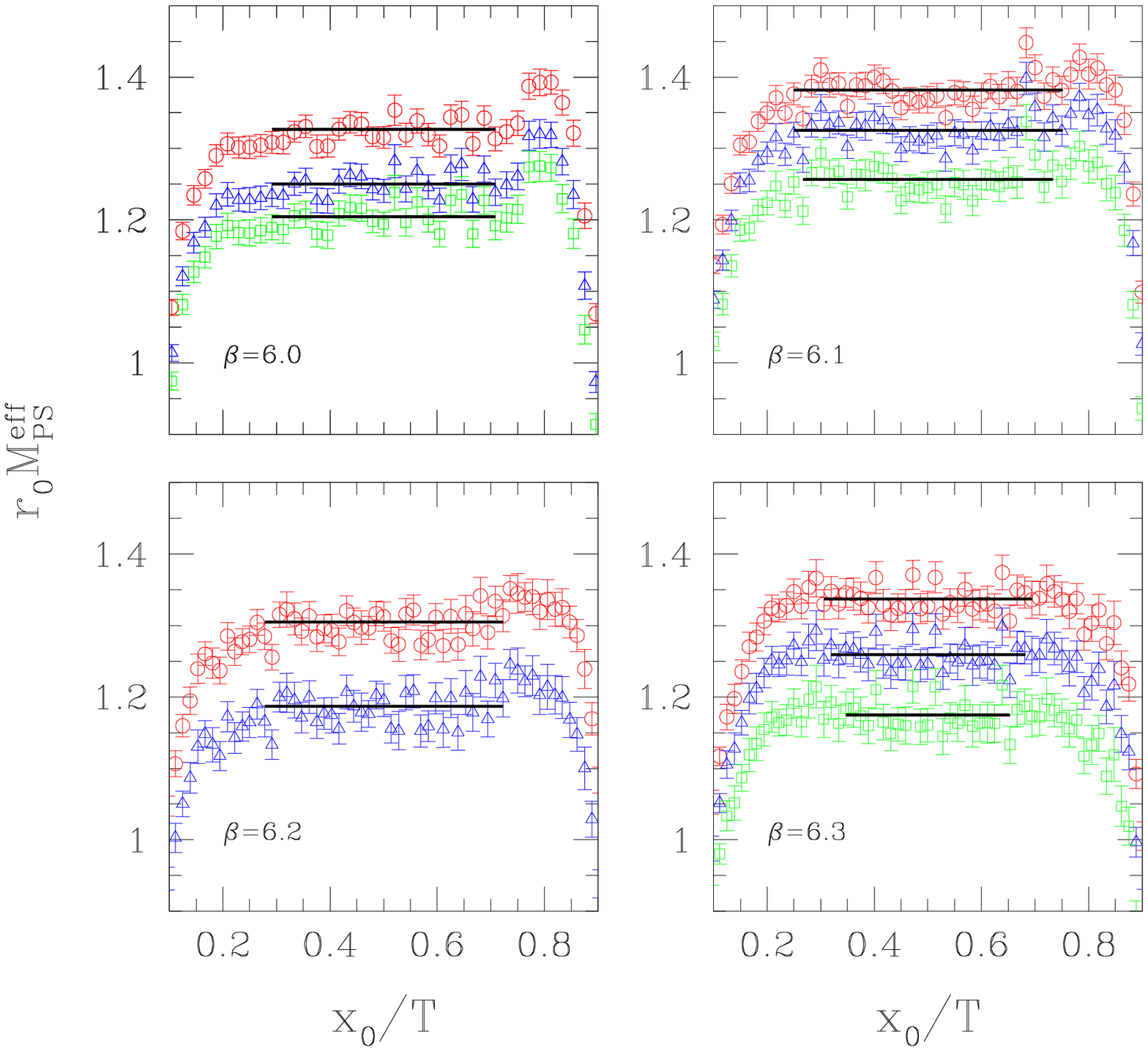, width=12.5 true cm}
\vskip -1.5cm
\epsfig{figure=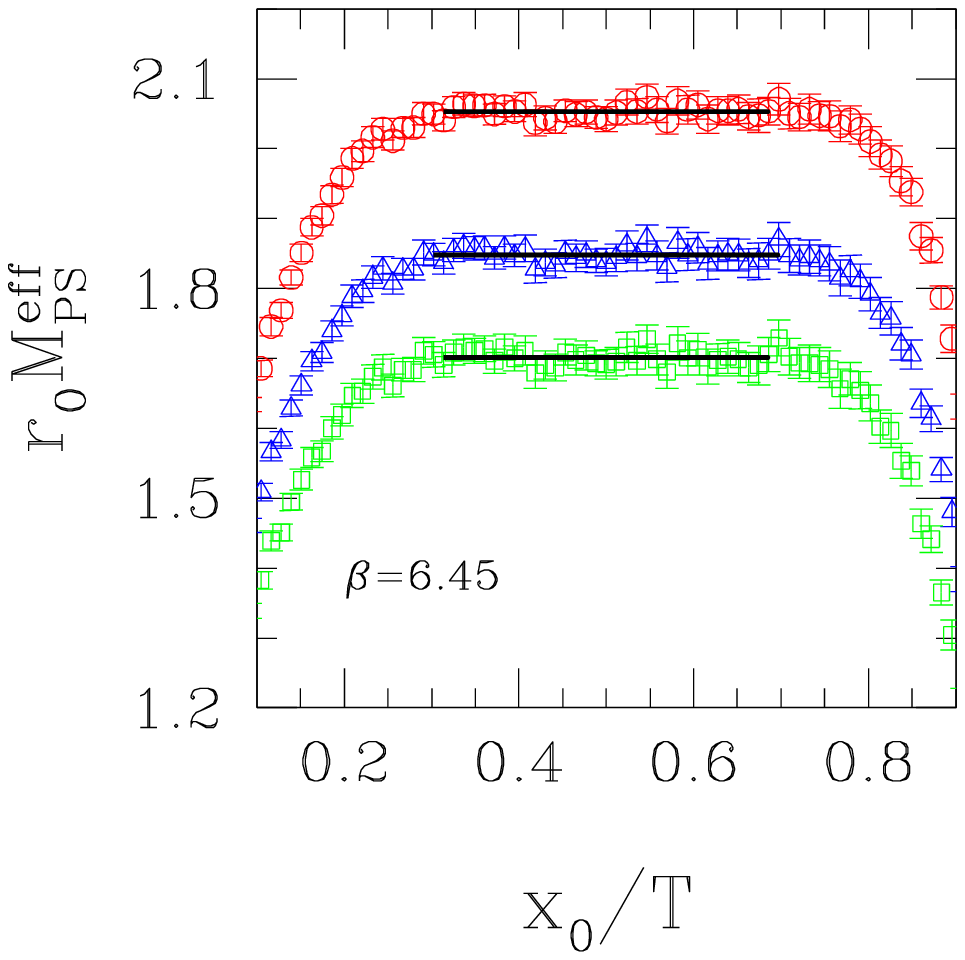, width=12.5 true cm}
\vskip -4.5cm
\caption{Plateaux for the extraction of $M_{\rm PS}^{\rm eff}$ at
$\alpha=\pi/4$. The time-range and value of each plateau is indicated
by a straight line segment.}
\label{fig:plateaux_m_pi4_1}
\end{center}
\end{figure}

\begin{figure}
\begin{center}
\epsfig{figure=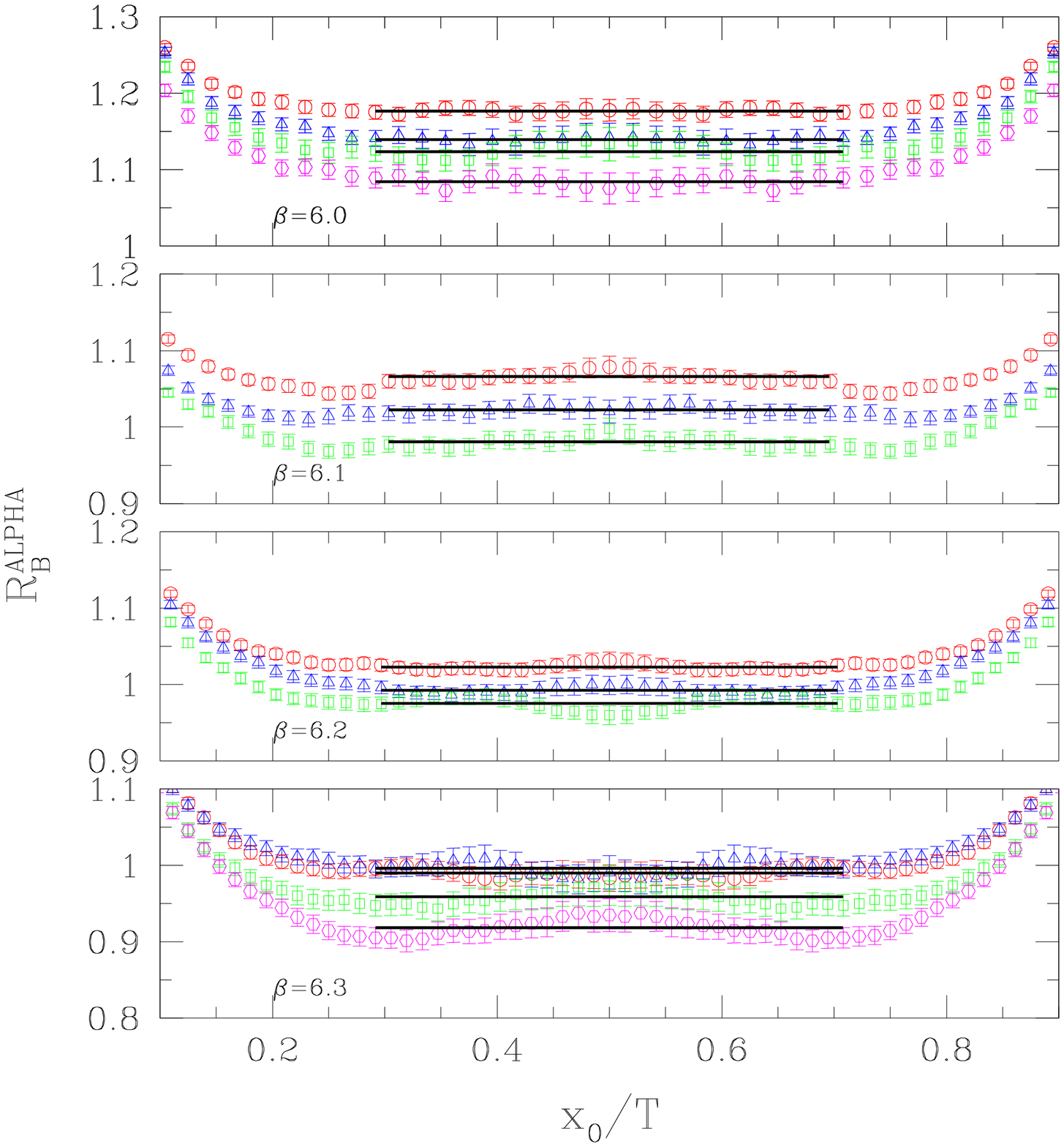, width=12.5 true cm}
\vskip -0.0cm
\caption{Plateaux for the extraction of $R_{\rm B}^{\rm ALPHA}$ at 
$\alpha=\pi/2$.  The time-range and value of each plateau is indicated
by a straight line segment.}
\end{center}
\end{figure}

\begin{figure}
\begin{center}
\epsfig{figure=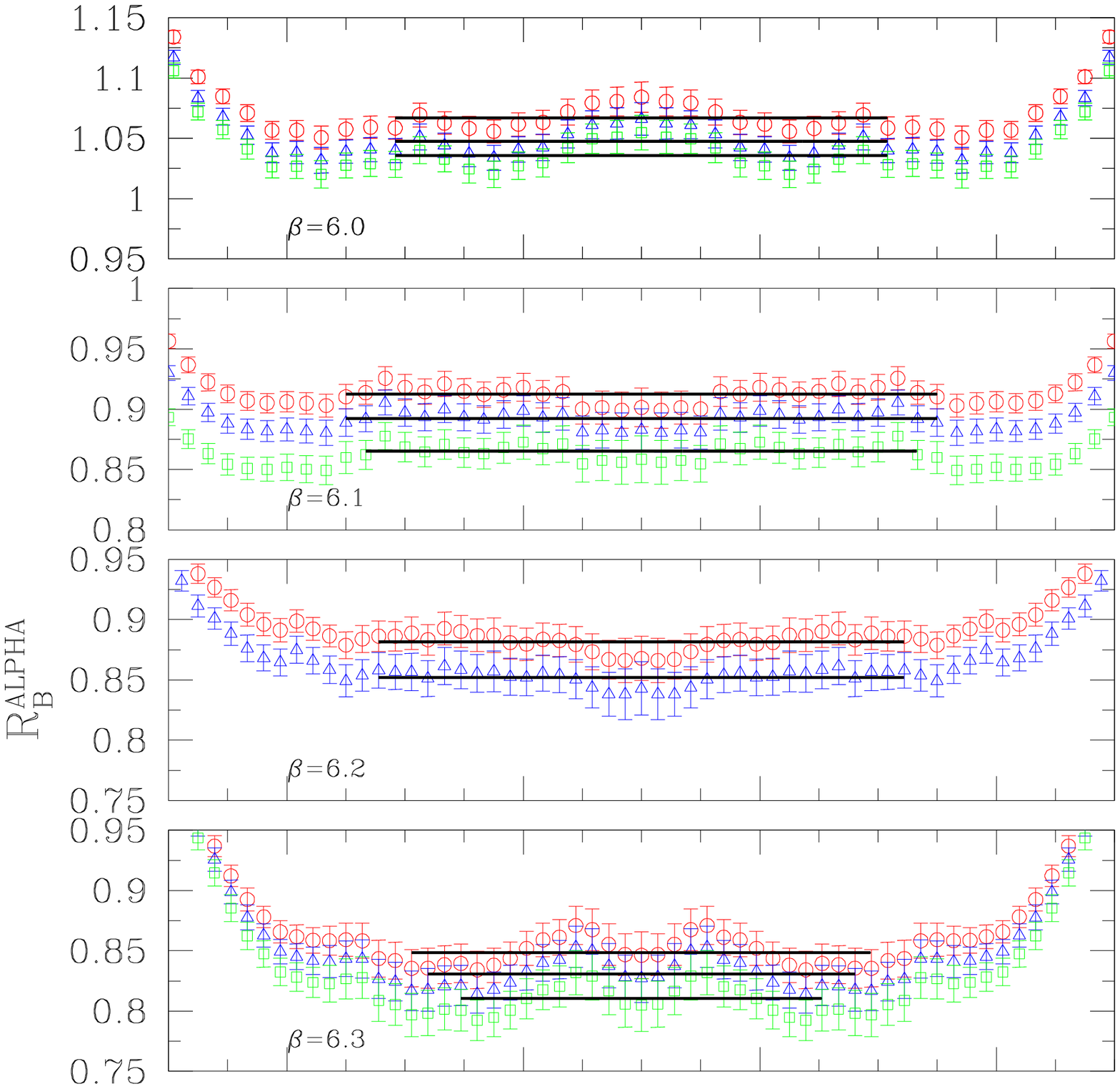, width=12.5 true cm}
\vskip -1.4cm
\epsfig{figure=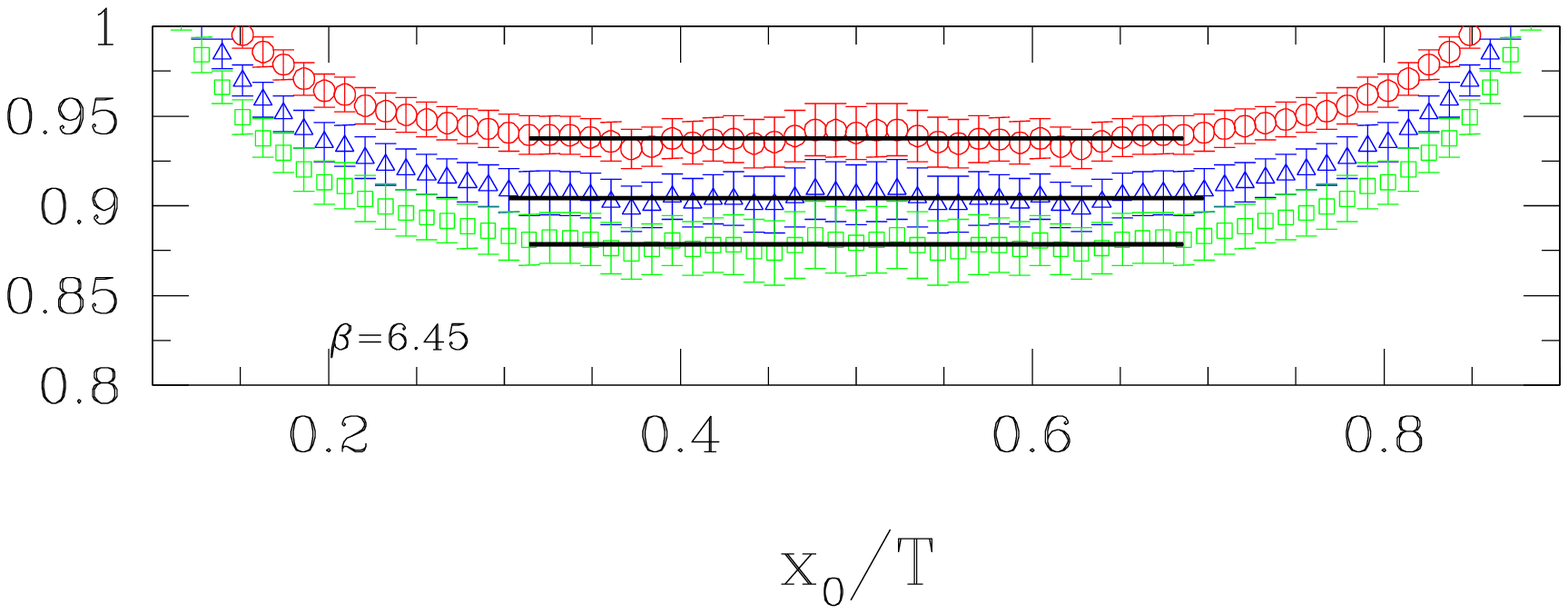, width=12.5 true cm}
\vskip -8.0cm
\caption{Plateaux for the extraction of $R_{\rm B}^{\rm ALPHA}$
at $\alpha=\pi/4$. The time-range and value of each plateau is indicated
by a straight line segment.}
\label{fig:plateaux_BK_Alpha_pi4}
\end{center}
\end{figure}

\begin{figure}
\begin{center}
\epsfig{figure=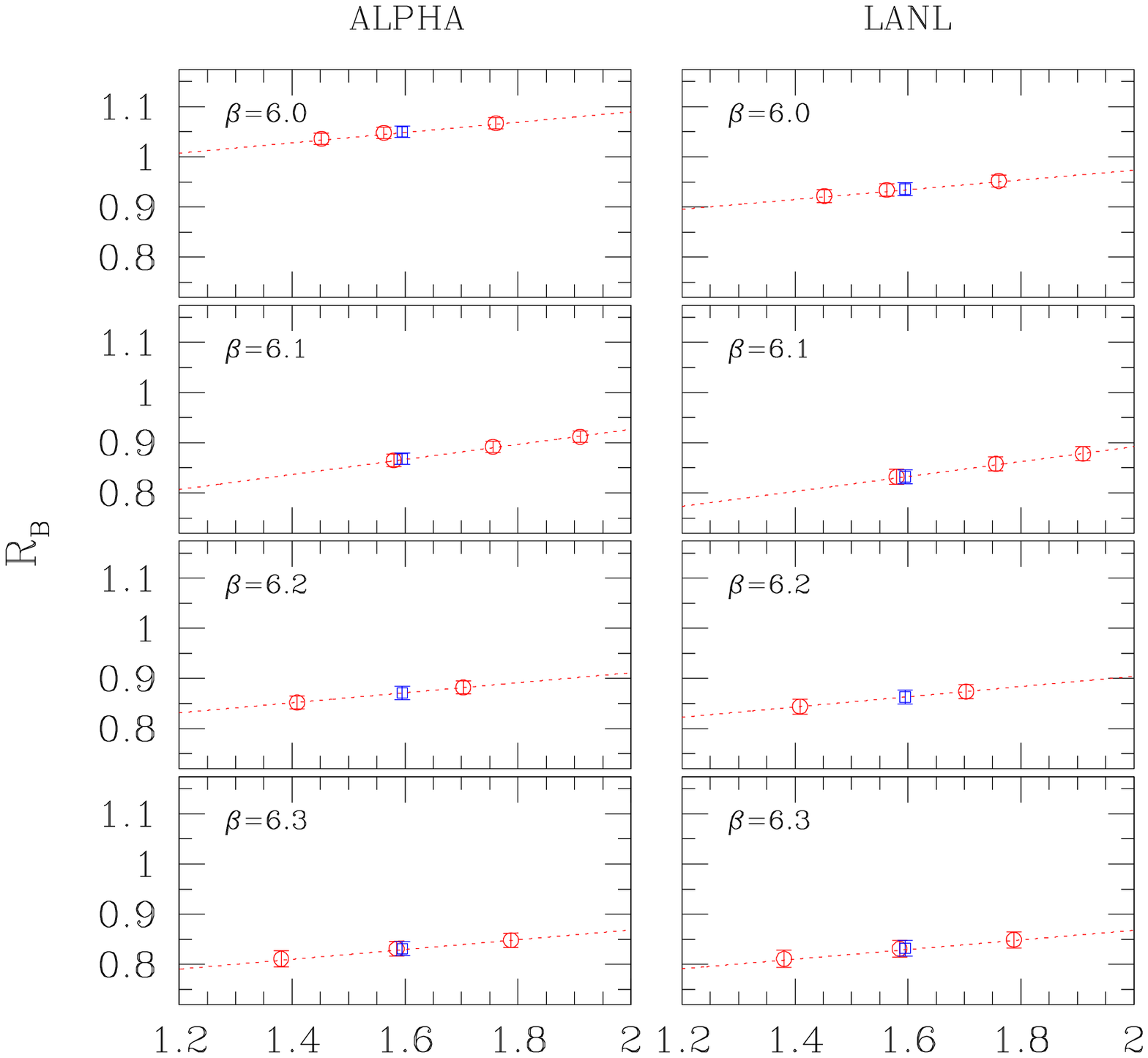, width=12.5 true cm}
\vskip -2.0cm
\epsfig{figure=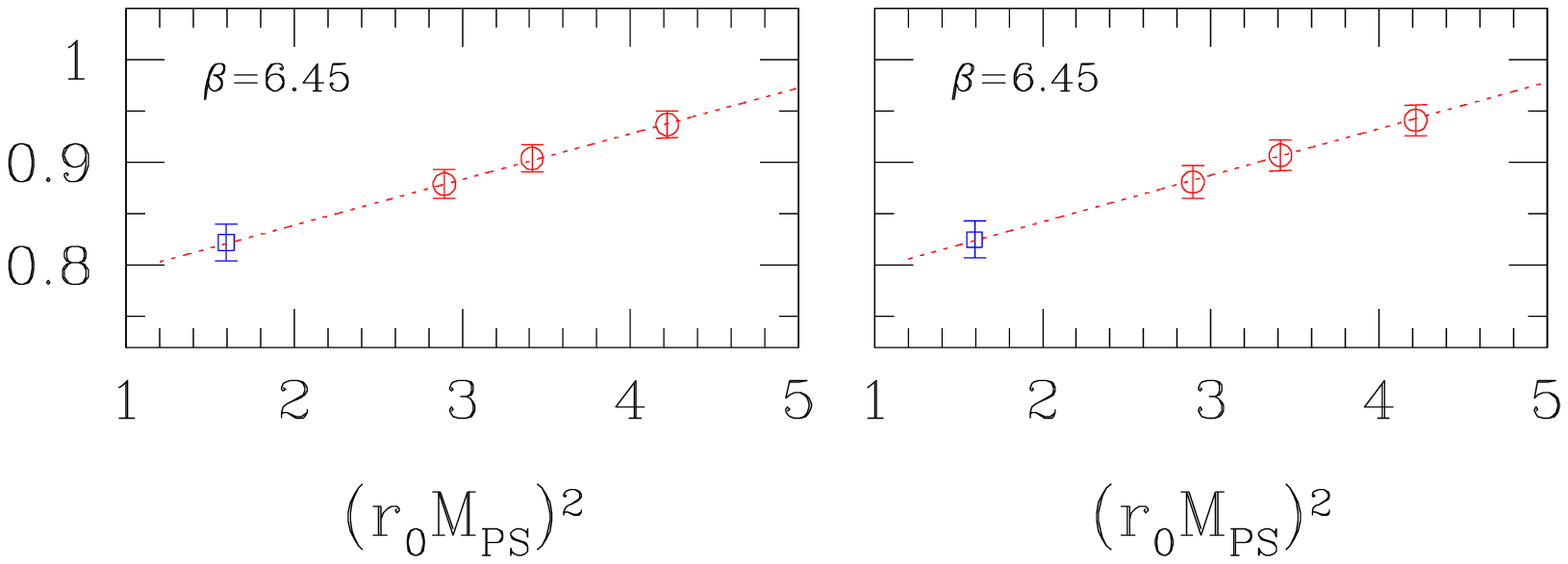, width=12.5 true cm}
\vskip -7.0cm
\caption{Linear inter/extrapolation of $R_{\rm B}$ to the physical
kaon mass at $\taa = \pi/4$.}
\label{fig:interp_pi4}
\end{center}
\end{figure}

\begin{figure}[!h]
\begin{center}
\epsfig{figure=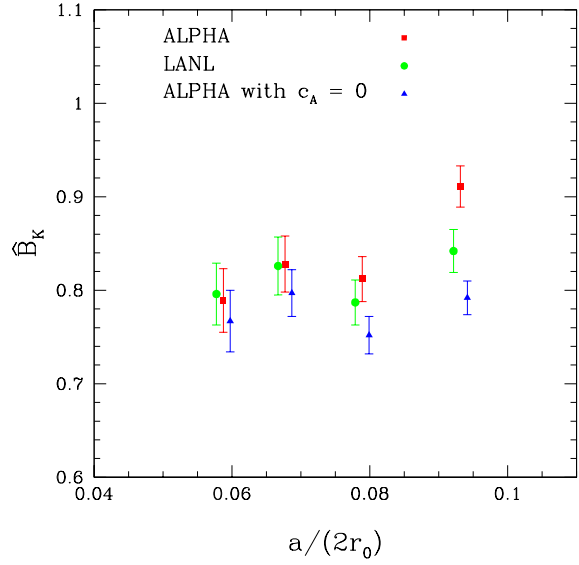, width=8.0 truecm}
\epsfig{figure=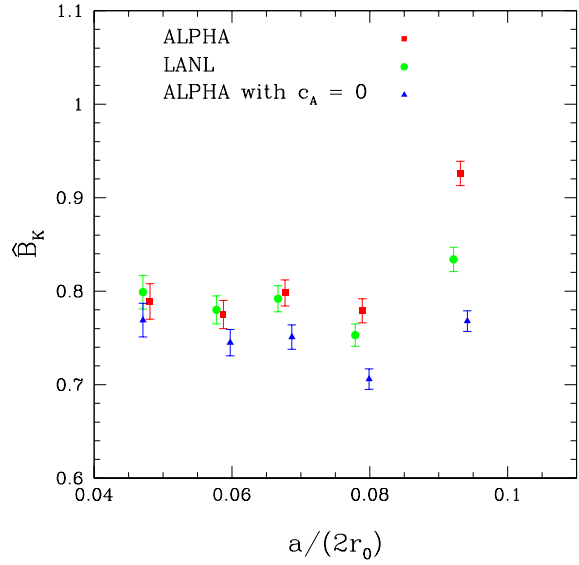, width=8.0 truecm}
\caption{Comparison of $\hat B_K$, extracted from three different
definitions of $R_{\rm B}$, differing by the axial currents'
$\Oa$ counterterms. Top: results for $\taa = \pi/2$.
Bottom: results for $\taa = \pi/4$.}
\label{fig:3R}
\end{center}
\end{figure}

\begin{figure}
\begin{center}
\epsfig{figure=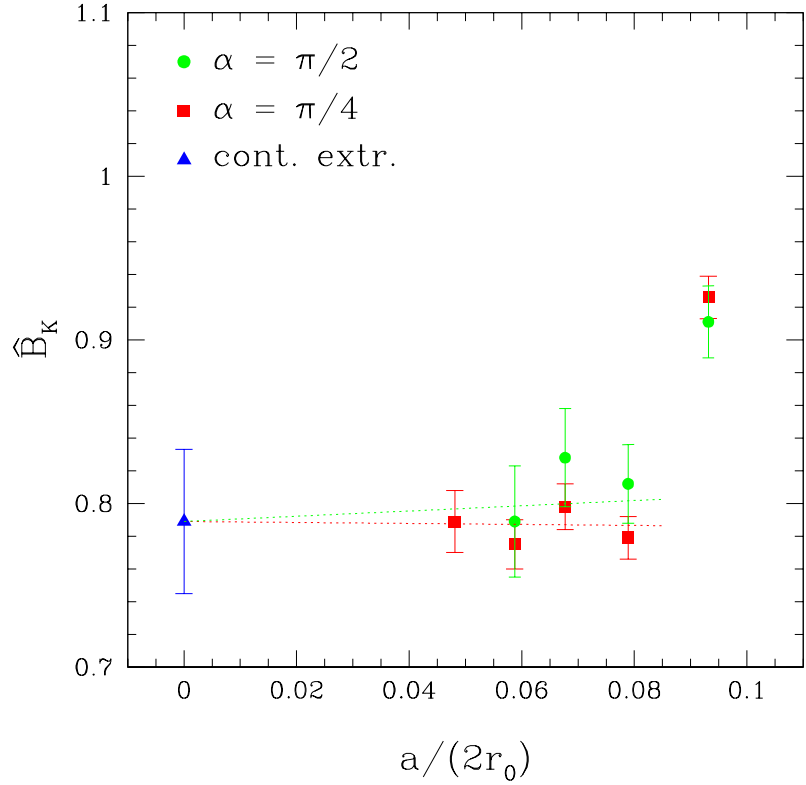, width=11.2 true cm}
\caption{Continuum limit extrapolation of $\hat B_K$, obtained
from a combined linear fit of the $\taa = \pi/2$ and $\taa = \pi/4$
data. The coarsest lattice data, corresponding to $\beta = 6.0$,
is not included in the fit.}
\label{fig:extrap_cl}
\end{center}
\end{figure}

\end{document}